\documentclass[aps,nofootinbib,prd,preprintnumbers,showpacs]{revtex4}
\usepackage{eurosym}
\usepackage{wrapfig}
\usepackage{amsfonts}
\usepackage{amsmath}
\usepackage{hyperref}
\usepackage{amssymb}
\usepackage[english]{babel}
\usepackage{graphicx}
\usepackage{epsfig}
\usepackage{bm}
\usepackage{longtable}
\usepackage{verbatim}
\usepackage{longtable}
\usepackage{color}
\usepackage{subfigure}
\usepackage[utf8]{inputenc}

\newcommand{\mathsym}[1]{{}}

\input{tcilatex}
\begin{document}

\title{A highly predictive $A_{4}$ flavour 3-3-1 model with radiative
inverse seesaw mechanism}
\author{A. E. C\'arcamo Hern\'andez${}^{a}$}
\email{antonio.carcamo@usm.cl}
\author{H. N. Long$^{b, c}$}
\email{Corresponding.author: hoangngoclong@tdt.edu.vn}
\affiliation{$^{{a}}$Universidad T\'{e}cnica Federico Santa Mar\'{\i}a and Centro Cient%
\'{\i}fico-Tecnol\'{o}gico de Valpara\'{\i}so, \\
Casilla 110-V, Valpara\'{\i}so, Chile,\\
$^{{b}}$Theoretical Particle Physics and Cosmology Research Group, Advanced Institute of Materials Science, Ton Duc
Thang University, Ho Chi Minh City, Vietnam\\
$^{{c}}$Faculty of Applied Sciences, Ton Duc Thang University, Ho Chi Minh
City, Vietnam}
\date{\today }

\begin{abstract}
We build a highly predictive 3-3-1 model, where the field content is
extended by including several $SU(3)_{L}$ scalar singlets and six right
handed Majorana neutrinos. In our model the $SU(3)_{C}\times SU\left(
3\right) _{L}\times U\left( 1\right) _{X}$ gauge symmetry is supplemented by
the $A_{4}\times Z_{4}\times Z_{6}\times Z_{16}\times Z_{16}^{\prime }$
discrete group, which allows to get a very good description of the low energy
fermion flavor data. In the model under consideration, the $A_{4}\times
Z_{4}\times Z_{6}\times Z_{16}\times Z_{16}^{\prime }$ discrete group is
broken at very high energy scale down to the preserved $Z_{2}$ discrete
symmetry, thus generating the observed pattern of SM fermion masses and
mixing angles and allowing the implementation of the loop level inverse
seesaw mechanism for the generation of the light active neutrino masses,
respectively. The obtained values for the physical observables in the quark
sector agree with the experimental data, whereas those ones for the lepton
sector also do, only for the case of inverted neutrino mass spectrum. The
normal neutrino mass hierarchy scenario of the model is ruled out by the
neutrino oscillation experimental data. We find 
an effective Majorana neutrino mass parameter of neutrinoless double beta decay of $%
m_{ee}=46.9$ meV, a leptonic Dirac CP violating phase of $-81.37^{\circ}$ and a Jarlskog invariant of about $10^{-2}$ for the inverted
neutrino mass hierarchy. The preserved $Z_{2}$ symmetry allows for a stable
scalar dark matter candidate.
\end{abstract}

\pacs{12.60.Cn,12.60.Fr,12.15.Lk,14.60.Pq}
\maketitle

\textbf{Keywords}: Extensions of electroweak gauge sector, Extensions of
electroweak Higgs sector, Electroweak radiative corrections, Neutrino mass
and mixing

\section{Introduction}

Despite its great consistency with the experimental data, the Standard Model
(SM) is unable to explain several issues such as, for example, the number of
fermion generations, the large hierarchy of fermion masses, the small quark
mixing angles and the sizeable leptonic mixing ones. Whereas in the quark
sector, the mixing angles are small, in the lepton sector two of the mixing
angles are large, and one mixing angle is small. Neutrino experiments have
brought clear evidence of neutrino oscillations from the measured neutrino
mass squared splittings. The three neutrino flavors mix and at least two of
the neutrinos have non vanishing masses, which according to neutrino
oscillation experimental data must be smaller than the SM charged fermion
masses by many orders of magnitude.

Models with an extended gauge symmetry are frequently used to tackle the
limitations of the SM. In particular, the models based on the gauge symmetry 
$SU(3)_{c}\otimes SU(3)_{L}\otimes U(1)_{X}$, also called 3-3-1 models, can
explain the origin of fermion generations thanks to the introduction of a
family non-universal $SU(3)_{L}$ symmetry \cite%
{Georgi:1978bv,Valle:1983dk,Pisano:1991ee,Foot:1992rh,Frampton:1992wt,Hoang:1996gi, Hoang:1995vq,Foot:1994ym,Boucenna:2015pav,Hernandez:2015ywg}%
, can provide an explanation for the origin of the family structure of the
fermions. These models have the following nice interesting features: 1) The
three family structure in the fermion sector naturally arises in the 3-3-1
models from the cancellation of chiral anomalies and asymptotic freedom in
QCD. 2) The fact that the third family is treated under a different
representation, can explain the large mass difference between the heaviest
quark family and the two lighter ones. 3) The 3-3-1 models allow the
quantization of electric charge \cite{deSousaPires:1998jc,VanDong:2005ux}.
4) These models have several sources of CP violation \cite%
{Montero:1998yw,Montero:2005yb}. 5) The above models explain why the
Weinberg mixing angle satisfies $\sin ^{2}\theta _{W}<\frac{1}{4}$. 6) These
models contain a natural Peccei-Quinn symmetry, necessary to solve the
strong-CP problem \cite{Pal:1994ba,Dias:2002gg,Dias:2003zt,Dias:2003iq}. 7)
The 3-3-1 models with heavy sterile neutrinos include cold dark matter
candidates as weakly interacting massive particles (WIMPs) \cite%
{Mizukoshi:2010ky,Dias:2010vt,Alvares:2012qv,Cogollo:2014jia}. A concise
review of WIMPs in 3-3-1 Electroweak Gauge Models is provided in Ref. \cite%
{daSilva:2014qba}.

In the 3-3-1 models, one heavy triplet field with a Vacuum Expectation Value
(VEV) at high energy scale $\nu _{\chi }$, breaks the symmetry $%
SU(3)_{L}\otimes U(1)_{X}$ into the SM electroweak group $SU(2)_{L}\otimes
U(1)_{Y}$, thus generating the masses of non SM fermions and non SM gauge
bosons, while the other two lighter triplets with VEVs at the electroweak
scale $\upsilon _{\rho }$ and $\upsilon _{\eta }$, trigger the Electroweak
Symmetry Breaking \cite{Hernandez:2013mcf} and provide the masses for the SM
particles.

On the other hand, the implementation of discrete flavor symmetries in
several extensions of the SM has provided a nice description of the observed
pattern of fermion masses and mixings (recent reviews on discrete flavor
groups can be found in Refs. \cite%
{Ishimori:2010au,Altarelli:2010gt,King:2013eh, King:2014nza}). Several
discrete groups have been employed in extensions of the SM, such as $A_{4}$ 
\cite%
{Ma:2001dn,He:2006dk,Chen:2009um,Ahn:2012tv,Memenga:2013vc,Felipe:2013vwa,Varzielas:2012ai, Ishimori:2012fg,King:2013hj,Hernandez:2013dta,Babu:2002dz,Altarelli:2005yx,Morisi:2013eca, Altarelli:2005yp,Kadosh:2010rm,Kadosh:2013nra,delAguila:2010vg,Campos:2014lla,Vien:2014pta,Hernandez:2015tna,CarcamoHernandez:2017cwi}%
, $S_{3}$ \cite%
{Kubo:2003pd,Kobayashi:2003fh,Chen:2004rr,Mondragon:2007af,Mondragon:2008gm,Bhattacharyya:2010hp, Dong:2011vb,Dias:2012bh,Meloni:2012ci,Canales:2012dr,Canales:2013cga,Ma:2013zca,Kajiyama:2013sza,Hernandez:2013hea, Ma:2014qra,Hernandez:2014vta,Hernandez:2014lpa,Gupta:2014nba,Hernandez:2015dga,Hernandez:2015zeh,Hernandez:2015hrt,Hernandez:2016rbi,CarcamoHernandez:2016pdu,Arbelaez:2016mhg}%
, $S_{4}$ \cite%
{Mohapatra:2012tb,BhupalDev:2012nm,Varzielas:2012pa,Ding:2013hpa,Ishimori:2010fs,Ding:2013eca,Hagedorn:2011un, Campos:2014zaa,Dong:2010zu,VanVien:2015xha,Arbelaez:2016mhg}%
, $D_{4}$ \cite%
{Frampton:1994rk,Grimus:2003kq,Grimus:2004rj,Frigerio:2004jg,Babu:2004tn,Adulpravitchai:2008yp, Ishimori:2008gp,Hagedorn:2010mq,Meloni:2011cc,Vien:2013zra}%
, $Q_{6}$ \cite%
{Kawashima:2009jv,Kaburaki:2010xc,Babu:2011mv,Gomez-Izquierdo:2013uaa,Gomez-Izquierdo:2017med}%
, $T_{7}$ \cite%
{Luhn:2007sy,Hagedorn:2008bc,Cao:2010mp,Luhn:2012bc,Kajiyama:2013lja,Bonilla:2014xla,Vien:2014gza, Vien:2015koa,Hernandez:2015cra,Arbelaez:2015toa}%
, $T_{13}$ \cite{Ding:2011qt,Hartmann:2011dn,Hartmann:2011pq,Kajiyama:2010sb}%
, $T^{\prime }$ \cite%
{Aranda:2000tm,Sen:2007vx,Aranda:2007dp,Chen:2007afa,Frampton:2008bz,Eby:2011ph,Frampton:2013lva,Chen:2013wba}%
, $\Delta (27)$ \cite%
{Ma:2007wu,Varzielas:2012nn,Bhattacharyya:2012pi,Ma:2013xqa,Nishi:2013jqa,Varzielas:2013sla,Aranda:2013gga,Ma:2014eka, Abbas:2014ewa,Abbas:2015zna,Varzielas:2015aua,Bjorkeroth:2015uou,Chen:2015jta,Vien:2016tmh,Hernandez:2016eod,CarcamoHernandez:2017owh,Bernal:2017xat}%
, $\Delta(54)$ \cite{Carballo-Perez:2016ooy} and $A_{5}$ \cite%
{Everett:2008et,Feruglio:2011qq,Cooper:2012bd,Varzielas:2013hga,Gehrlein:2014wda,Gehrlein:2015dxa,DiIura:2015kfa,Ballett:2015wia, Gehrlein:2015dza,Turner:2015uta,Li:2015jxa}
have been considered to explain the observed pattern of fermion masses and
mixings.

Among several discrete symmetry groups, the $A_{4}$ group has attracted a
lot of attention since it is the smallest one which admits one
three-dimensional representation as well as three inequivalent
one-dimensional representations. Then, the choice of the $A_{4}$ symmetry is
natural since there are three families of fermions, i.e, the left handed
leptons can be unified in triplet representation of $A_{4}$ while the right
handed leptons can be assigned to $A_{4}$ singlets. This setup has been
proposed for first time in Ref. \cite{Ma:2001dn} to study the lepton masses
and mixings obtaining nearly degenerate neutrino masses and allowing
realistic charged leptons masses after the $A_{4}$ symmetry is spontaneously
broken. The scalar sector of the minimal setup of Ref. \cite{Ma:2001dn}
includes one $A_{4}$ triplet whose components are $SU(2)_{L}$ doublets and
one $SU(2)_{L}$ doublet which transforms as an $A_{4}$ trivial singlet. As
it has been extensively discussed in the literature (for a recent reviews
see Refs. \cite{King:2013eh,Altarelli:2010gt,Ishimori:2010au}) the $A_{4}$
group, which is the group of even permutations of four elements has been
shown to generate the tribimaximal mixing pattern which predicts solar
mixing and atmospheric mixing angles consistent with the experimental data
but yields a vanishing reactor mixing angle contradicting the recent
experimental results 
from the Daya Bay \cite{An:2012eh}, T2K \cite{Abe:2011sj}, MINOS \cite%
{Adamson:2011qu}, Double CHOOZ \cite{Abe:2011fz} and RENO \cite{Ahn:2012nd}
experiments. In view of this the tribimaximal mixing pattern has to be
modified.

In this work we build a highly predictive $A_{4}$ flavor 3-3-1 model, where
the $A_{4}$ discrete symmetry is supplemented by the $Z_{4}\times
Z_{6}\times Z_{16}\times Z_{16}^{\prime }$ discrete group, providing a
framework consistent with the current low energy fermion flavor data. In the
model under consideration the different discrete group factors are broken
completely, excepting the $Z_{6}$ discrete group, which is broken down to
the preserved $Z_{2}$ symmetry, thus allowing the implementation of the one
loop level inverse seesaw mechanism for the generation of the light active
neutrino masses. The SM charged fermion masses and quark mixing angles arise
from the breaking of the $A_{4}\times Z_{4}\times Z_{6}\times Z_{16}\times
Z_{16}^{\prime }$ discrete group. 

%
The content of this paper goes as follows. In section \ref{model} we
describe our model. The low energy scalar potential of our model is
discussed in Section \ref{scalarpotential}. Section \ref{quarksector} is
devoted to the implications of our model in quark masses and mixings.
Section \ref{leptonsector} deals with lepton masses and mixings. We conclude
in section \ref{conclusions}. Appendix \ref{A4} provides a concise
description of the $A_{4}$ discrete group. Appendix \ref{B1} shows a
discussion of the scalar potential for a $A_{4}$ scalar triplet and its
minimization equations.

\section{The model.}

\label{model}

As is well known, the $SU(3)_{C}\times SU\left( 3\right) _{L}\times U\left(
1\right) _{X}$ model (3-3-1 model) with $\beta =-\frac{1}{\sqrt{3}}$ and
right-handed Majorana neutrinos in the $SU(3)_{L}$ lepton triplet is
unsatisfactory in describing the observed SM fermion mass and mixing
pattern, due to the unexplained hierarchy among its large number of Yukawa
couplings. To address that problem, we propose an extension of the 3-3-1
model with $\beta =-\frac{1}{\sqrt{3}}$, where the scalar sector is extended
to include several EW scalar singlets, the fermion sector is extended by
introducing six right handed Majorana neutrinos, and the $SU(3)_{C}\times
SU\left( 3\right) _{L}\times U\left( 1\right) _{X}$ \ gauge symmetry is
supplemented by the $A_{4}\times Z_{4}\times Z_{6}\times Z_{16}\times
Z_{16}^{\prime }$ discrete group, so that the full symmetry $\mathcal{G}$
exhibits the following three-step spontaneous breaking: 
\begin{eqnarray}
&&\mathcal{G}=SU(3)_{C}\times SU\left( 3\right) _{L}\times U\left( 1\right)
_{X}\times A_{4}\times Z_{4}\times Z_{6}\times Z_{16}\times Z_{16}^{\prime }
\label{Group} \\
&&\hspace{35mm}\Downarrow \Lambda _{int}  \notag \\[0.12in]
&&\hspace{15mm}SU(3)_{C}\times SU\left( 3\right) _{L}\times U\left( 1\right)
_{X}\times Z_{2}\times Z_{4}  \notag \\[0.12in]
&&\hspace{35mm}\Downarrow v_{\chi }  \notag \\[0.12in]
&&\hspace{15mm}SU(3)_{C}\times SU\left( 2\right) _{L}\times U\left(
1\right) _{Y}\times Z_{2}  \notag \\[0.12in]
&&\hspace{35mm}\Downarrow v_{\eta },v_{\rho }  \notag \\[0.12in]
&&\hspace{15mm}SU(3)_{C}\times U\left( 1\right) _{Q}\otimes Z_{2}  \notag
\end{eqnarray}%
where the different symmetry breaking scales satisfy the following hierarchy 
$\Lambda _{int}\gg v_{\chi }\gg v_{\eta },v_{\rho }.$ Let us note that all
discrete group are broken completely at the very high energy scale $\Lambda
_{int}\gg v_{\chi }$, excepting the $Z_{6}$ discrete group which is broken
down to the preserved $Z_{2}$ symmetry. That preserved \ $Z_{2}$ symmetry
will allows us to implement a one loop level inverse seesaw mechanism for
the generation of the light active neutrino masses.

In the 3-3-1 model under consideration, the electric charge is defined in
terms of the $SU(3)$ generators and the identity by: 
\begin{equation}
Q=T_{3}+\beta T_{8}+XI=T_{3}-\frac{1}{\sqrt{3}}T_{8}+XI,
\end{equation}%
with $I=diag(1,1,1)$, $T_{3}=\frac{1}{2}diag(1,-1,0)$ and $T_{8}=(\frac{1}{2%
\sqrt{3}})diag(1,1,-2)$ for triplet. Let us note that we have chosen $\beta
=-\frac{1}{\sqrt{3}}$, because in that choice the third component of the
weak lepton triplet is a neutral field $\nu _{R}^{C}$ which allows to build
the Dirac matrix with the usual field $\nu _{L}$ of the weak doublet. The
introduction of a sterile neutrino $N_{R}$ in the model allows the
implementation of a low scale seesaw mechanism (which could be inverse or
linear) for the generation of the light neutrino masses. The 3-3-1 models
with $\beta =-\frac{1}{\sqrt{3}}$ have the advantage over other 3-3-1 models
with different values $\beta $, of providing an alternative framework to
generate neutrino masses, where the neutrino spectrum includes the light
active sub-eV scale neutrinos as well as sterile neutrinos which could be
dark matter candidates if they are light enough or candidates for detection
at the LHC, if they have TeV scale masses. Let us note that if the TeV scale
sterile neutrinos are found at the LHC, the 3-3-1 models with $\beta =-\frac{%
1}{\sqrt{3}}$ can be very strong candidates for unraveling the mechanism
responsible for electroweak symmetry breaking.

The cancellation of chiral anomalies implies that quarks are unified in the
following $SU(3)_{C}\times SU(3)_{L}\times U(1)_{X}$ left- and right-handed
representations \cite%
{Valle:1983dk,Hoang:1995vq,Diaz:2004fs,CarcamoHernandez:2005ka}: 
\begin{align}
Q_{nL}& =%
\begin{pmatrix}
D_{n} \\ 
-U_{n} \\ 
J_{n} \\ 
\end{pmatrix}%
_{L}\sim \left( 3,3^{\ast },0\right) ,\hspace{1cm}Q_{3L}=%
\begin{pmatrix}
U_{3} \\ 
D_{3} \\ 
T \\ 
\end{pmatrix}%
_{L}\sim \left( 3,3,\frac{1}{3}\right) ,\hspace{1cm}n=1,2,  \notag \\
D_{iR}& \sim \left( 3,1,-\frac{1}{3}\right) ,\hspace{1cm}U_{iR}\sim \left(
3,1,\frac{2}{3}\right) ,\hspace{1cm}J_{nR}\sim \left( 3,1,-\frac{1}{3}%
\right) ,\hspace{1cm}T_{R}\sim \left( 3,1,\frac{2}{3}\right) ,\hspace{1cm}%
i=1,2,3,
\end{align}
where $U_{iL}$ and $D_{iL}$ ($i=1,2,3$) are the left handed up and down type
quarks fields in the flavor basis, respectively. The right handed SM\
quarks, i.e., $U_{iR}$ and $D_{iR}$ ($i=1,2,3$) and right handed exotic
quarks, i.e., $T_{R}$ and $J_{nR}$ ($n=1,2$)\ are assigned as $SU(3)_{L}$ singlets with $U(1)_{X}$ quantum numbers equal to their electric charges.

Furthermore, the requirement of chiral anomaly cancellation constrains the
leptons to the following $SU(3)_{C}\times SU(3)_{L}\times U(1)_{X}$ left-
and right-handed representations \cite{Valle:1983dk,Hoang:1995vq,Diaz:2004fs}%
: 
\begin{equation}
L_{iL}=%
\begin{pmatrix}
\nu _{i} \\ 
e_{i} \\ 
\nu _{i}^{c} \\ 
\end{pmatrix}%
_{L}\sim \left( 1,3,-\frac{1}{3}\right) ,\hspace{1cm}e_{iR}\sim \left(
1,1,-1\right) ,\hspace{1cm}i=1,2,3,  \label{L}
\end{equation}

In the present model the fermion sector is extended by introducing six right
handed Majorana neutrinos, singlets under the 3-3-1 group, so that they have
the following $SU(3)_{C}\times SU(3)_{L}\times U(1)_{X}$ assignments: 
\begin{equation}
N_{iR}\sim \left( 1,1,0\right) ,\hspace{1cm}\Omega _{iR}\sim \left(
1,1,0\right) ,\hspace{1cm}i=1,2,3,
\end{equation}

Regarding the scalar sector of the 3-3-1 model with right handed Majorana
neutrinos, we assign the scalar fields in the following $SU(3)_{C}\times
SU(3)_{L}\times U(1)_{X}$ representations: 
\begin{align}
\chi & =%
\begin{pmatrix}
\chi _{1}^{0} \\ 
\chi _{2}^{-} \\ 
\frac{1}{\sqrt{2}}(v_{\chi }+\xi _{\chi }\pm i\zeta _{\chi })%
\end{pmatrix}%
\sim \left( 1,3,-\frac{1}{3}\right) ,\hspace{1cm}\rho =%
\begin{pmatrix}
\rho _{1}^{+} \\ 
\frac{1}{\sqrt{2}}(v_{\rho }+\xi _{\rho }\pm i\zeta _{\rho }) \\ 
\rho _{3}^{+}%
\end{pmatrix}%
\sim \left( 1,3,\frac{2}{3}\right) ,  \notag \\
\eta & =%
\begin{pmatrix}
\frac{1}{\sqrt{2}}(v_{\eta }+\xi _{\eta }\pm i\zeta _{\eta }) \\ 
\eta _{2}^{-} \\ 
\eta _{3}^{0}%
\end{pmatrix}%
\sim \left( 1,3,-\frac{1}{3}\right) ,  
\end{align}

\quad The scalar sector of the 3-3-1 model with right handed Majorana
neutrinos includes: three $3$'s irreps of $SU(3)_{L}$, where one triplet $%
\chi $ gets a TeV scale vacuum expectation value (VEV) $v_{\chi }$, that
breaks the $SU(3)_{L}\otimes U(1)_{X}$ symmetry down to $SU(2)_{L}\otimes
U(1)_{Y}$, thus generating the masses of non SM fermions and non SM gauge
bosons; and two light triplets $\eta $ and $\rho $ acquiring electroweak
scale VEVs $v_{\eta }$ and $v_{\rho }$, respectively, thus triggering
Electroweak Symmetry Breaking and then providing masses for the fermions and
gauge bosons of the SM \cite{Hernandez:2013mcf}.

On the other hand, the lepton number has a gauge component as well as a complementary global one, according to the following relation:
\begin{equation}
L=\frac{4}{\sqrt{3}}T_{8}+\mathcal{L}=\left( 
\begin{array}{ccc}
\pm\frac{2}{3}+\mathcal{L} & 0 & 0 \\ 
0 & \pm\frac{2}{3}+\mathcal{L} & 0 \\ 
0 & 0 & \mp\frac{4}{3}+\mathcal{L} 
\end{array}%
\right),
\label{Leptonnumber}
\end{equation}%
where the upper and lower signs correspond to triplet and antitriplet of $SU(3)_L$, respectively. The $L$ operator that does not commute with the $SU(3)_C\times SU(3)_L\times U(1)_X$ gauge symmetry. However, $\mathcal{L}$ is a conserved charge corresponding to the $U(1)_{\mathcal{L}}$ global symmetry, commuting with the gauge symmetry and corresponds to the ordinary lepton number. In addition, the hypercharge operator is defined as:
\begin{equation}
\frac{Y}{2}=-\frac1{4}\left(L-\mathcal{L}\right)+X 
\end{equation}
From Eq. (\ref{Leptonnumber}) it follows that the masses of the right handed Majorana neutrinos $N_{jR}$, which will be generated at one loop level (as we will shown later, in section \ref{leptonsector}) will break the lepton number by two units, thus allowing the implementation of the one loop level inverse seesaw mechanism for the generation of the light active neutrino masses and giving rise to the neutrinoless double beta decay. Consequently, the light active neutrinos are also Majorana particles, as follows from the Valle-Schechter Theorem \cite{Schechter:1981bd}, which states that any mechanism generating neutrinoless double beta decay implies that neutrinos are Majorana particles.

We extend the scalar sector of the 3-3-1 model with right handed Majorana
neutrinos by adding the following $SU(3)_{L}$ scalar singlets, with the
following $SU(3)_{C}\times SU(3)_{L}\times U(1)_{X}$ assignments: 
\begin{align}
\varphi & \sim \left( 1,0,0\right) ,\hspace{0.7cm}\tau _{n}\sim \left(
1,0,0\right) ,\hspace{0.7cm}\sigma \sim \left( 1,0,0\right) ,\hspace{0.7cm}%
\phi \sim \left( 1,0,0\right) ,\hspace{0.7cm}n=1,2,  \notag \\
\varrho & \sim \left( 1,0,0\right) ,\hspace{0.7cm}\xi _{j}\sim \left(
1,0,0\right) ,\hspace{0.7cm}\zeta _{j}\sim \left( 1,0,0\right) ,\hspace{0.7cm%
}\Phi _{j}\sim \left( 1,0,0\right) ,\hspace{0.7cm}\Delta _{j}\sim \left(
1,0,0\right) ,  \notag \\
\Xi _{j}& \sim \left( 1,0,0\right) ,\hspace{0.7cm}\Theta _{j}\sim \left(
1,0,0\right) ,\hspace{0.7cm}j=1,2,3.  \label{SU3Lscalarsinglets}
\end{align}

The scalar fields of our model have the following $A_{4}\times Z_{4}\times
Z_{6}\times Z_{16}\times Z_{16}^{\prime }$ assignments: 
\begin{eqnarray}
\chi &\sim &\left( \mathbf{1,}0,0,0,0\right) ,\hspace{1.5cm}\rho \sim \left( 
\mathbf{1,}0,2,0,0\right) ,\hspace{1.5cm}\eta \sim \left( \mathbf{1,}%
0,4,0,0\right) ,\hspace{1.5cm}\varphi \sim \left( \mathbf{1,}1,-3,0,0\right)
,  \notag \\
\sigma &\sim &\left( \mathbf{\mathbf{1}^{\prime \prime },}0,0,-1,0\right) ,%
\hspace{1.5cm}\tau _{1}\sim \left( \mathbf{1}^{\prime }\mathbf{,}%
0,0,-1,-1\right) ,\hspace{1.5cm}\tau _{2}\sim \left( \mathbf{1}^{\prime }%
\mathbf{,}0,0,-2,-1\right) ,\hspace{1.5cm}\phi \sim \left( \mathbf{1,}%
0,0,-1,0\right) ,  \notag \\
\varrho &\sim &\left( \mathbf{1,}2,0,0,0\right) ,\hspace{1.5cm}\xi \sim
\left( \mathbf{3,}0,0,6,-1\right) ,\hspace{1.5cm}\zeta \sim \left( \mathbf{3,%
}0,2,0,0\right) ,\hspace{1.5cm}\Phi \sim \left( \mathbf{3,}0,0,0,-3\right) ,
\notag \\
\Delta &\sim &\left( \mathbf{3,}0,0,0,-3\right) ,\hspace{1.5cm}\Xi \sim
\left( \mathbf{3,}0,0,0,-8\right) ,\hspace{1.5cm}\Theta \sim \left( \mathbf{%
3,}2,0,0,0\right) .
\end{eqnarray}%
Here the dimensions of the $A_{4}$ irreducible representations are specified
by the numbers in boldface and the different $Z_{4}\times Z_{6}\times
Z_{16}\times Z_{16}^{\prime }$ charges are written in additive notation. Let
us note that all scalar fields acquire nonvanishing vacuum expectation
values, excepting the $SU(3)_{L}$ scalar singlet $\varphi $, whose $Z_{6}$
charge corresponds to a nontrivial charge under the preserved $Z_{2}$
symmetry. The scalar assignments under the $U(1)_{\mathcal{L}}\times A_{4}\times Z_{4}\times
Z_{6}\times Z_{16}\times Z_{16}^{\prime }$ discrete group are summarized in
Table \ref{ta:scalars}.

\begin{table}[tbp]
\begin{tabular}{|c|c|c|c|c|c|c|c|c|c|c|c|c|c|c|c|}
\hline
& $\chi$ & $\rho$ & $\eta$ & $\varphi$ & $\sigma$ & $\tau_1$ & $\tau_2$ & $%
\phi$ & $\varrho$ & $\xi$ & $\zeta$ & $\Phi$ & $\Delta$ & $\Xi$ & $\Theta$
\\ \hline
$\mathcal{L}$ & $\frac{4}{3}$ & $-\frac{2}{3}$ & $-\frac{2}{3}$ & $2$ & $0$ & $0$ & $0$
& $0$ & $2$ & $0$ & $0$ & $0$ & $0$ & $0$ & $2$ \\ \hline
$A_{4}$ & $\mathbf{1}$ & $\mathbf{1}$ & $\mathbf{1}$ & $\mathbf{1}$ & $%
\mathbf{1}^{\prime \prime}$ & $\mathbf{1}^{\prime}$ & $\mathbf{1}^{\prime}$
& $\mathbf{1}$ & $\mathbf{1}$ & $\mathbf{3}$ & $\mathbf{3}$ & $\mathbf{3}$ & 
$\mathbf{3}$ & $\mathbf{3}$ & $\mathbf{3}$ \\ \hline
$Z_{4}$ & $0$ & $0$ & $0$ & $1$ & $0$ & $0$ & $0$ & $0$ & $2$ & $0$ & $0$ & $%
0$ & $0$ & $0$ & $2$ \\ \hline
$Z_{6}$ & $0$ & $2$ & $4$ & $-3$ & $0$ & $0$ & $0$ & $0$ & $0$ & $0$ & $2$ & 
$0$ & $0$ & $0$ & $0$ \\ \hline
$Z_{16}$ & $0$ & $0$ & $0$ & $0$ & $-1$ & $-1$ & $-2$ & $-1$ & $0$ & $6$ & $%
0 $ & $0 $ & $0$ & $0$ & $0$ \\ \hline
$Z_{16}^{\prime}$ & $0$ & $0$ & $0$ & $0$ & $0$ & $-1$ & $-1$ & $0$ & $0$ & $%
-1$ & $0$ & $-3$ & $-3$ & $-8$ & $0$ \\ \hline
\end{tabular}%
\caption{Scalar assignments under $U(1)_{\mathcal{L}}\times A_{4}\times Z_4\times Z_{6}\times
Z_{16}\times Z_{16}^{\prime }$.}
\label{ta:scalars}
\end{table}

The quark assignments under the group $A_{4}\times Z_{4}\times Z_{6}\times
Z_{16}\times Z_{16}^{\prime }$ are: 
\begin{eqnarray}
Q_{1L} &\sim &\left( \mathbf{1}^{\prime \prime }\mathbf{,}0,0,0,0\right) ,%
\hspace{1.5cm}Q_{2L}\sim \left( \mathbf{1}^{\prime },0,0,2,0\right) ,\hspace{%
1.5cm}Q_{3L}\sim \left( \mathbf{1,}0,0,4,0\right) ,  \notag \\
U_{1R} &\sim &\left( \mathbf{1,}0,4,8,8\right) ,\hspace{1.5cm}U_{2R}\sim
\left( \mathbf{1,}0,4,6,4\right) ,\hspace{1.5cm}U_{3R}\sim \left( \mathbf{1,}%
0,4,4,0\right) ,\hspace{1.5cm}T_{R}\sim \left( \mathbf{1,}0,0,4,0\right) , 
\notag \\
D_{R} &=&\left( D_{1R},D_{2R},D_{3R}\right) \sim \left( \mathbf{3,}%
0,2,0,1\right) ,\hspace{1.5cm}J_{1R}\sim \left( \mathbf{1}^{\prime \prime }%
\mathbf{,}0,0,0,0\right) ,\hspace{1.5cm}J_{2R}\sim \left( \mathbf{1}^{\prime
},0,0,2,0\right) .
\end{eqnarray}%
Lets us note that we assign the quarks fields into $A_{4}$ singlet
representations, excepting the SM right handed down type quarks fields which
are grouped in a $A_{4}$ triplet.

The lepton fields of our model have the following $A_{4}\times Z_{4}\times
Z_{6}\times Z_{16}\times Z_{16}^{\prime }$ assignments: 
\begin{eqnarray}
L_{L} &=&\left( L_{1L},L_{2L},L_{3L}\right) \sim \left( \mathbf{3,}%
0,3,0,0\right) ,\hspace{1cm}N_{R}=\left( N_{1R},N_{2R},N_{3R}\right) \sim
\left( \mathbf{3,}0,3,0,0\right), \\
\Omega _{R}&=&\left(\Omega_{1R},\Omega_{2R},\Omega _{3R}\right) \sim \left( 
\mathbf{3,}-1,0,0,0\right),\hspace{1cm}e_{1R}\sim\left( \mathbf{1,}%
0,3,5,3\right),\hspace{1cm}e_{2R}\sim \left( \mathbf{1,}0,3,0,8\right) ,%
\hspace{1cm}e_{3R}\sim \left(\mathbf{1,}0,3,0,3\right)  \notag
\end{eqnarray}%
As regards the lepton sector, we recall that the left and right-handed
leptons are grouped into $A_{4}$ triplet and $A_{4}$ singlet irreducible
representations, respectively, whereas the right-handed Majorana neutrinos,
i.e., $N_{iR}$ and are unified $\Omega _{iR}$ ($i=1,2,3$)\ into the $A_{4}$
triplets, i.e., $N_{R}$ and $\Omega _{R}$. The fermion assignments under the 
$U(1)_{\mathcal{L}}\times A_{4}\times Z_{4}\times Z_{6}\times Z_{16}\times Z_{16}^{\prime }$ discrete
group are summarized in Table \ref{ta:fermions}. 
\begin{table}[tbp]
\begin{tabular}{|c|c|c|c|c|c|c|c|c|c|c|c|c|c|c|c|c|}
\hline
& $Q_{1L}$ & $Q_{2L}$ & $Q_{3L}$ & $U_{1R}$ & $U_{2R}$ & $U_{3R}$ & $T_R$ & $%
D_R$ & $J_{1R}$ & $J_{2R}$ & $L_L$ & $N_R$ & $\Omega$ & $e_{1R}$ & $e_{2R}$
& $e_{3R}$ \\ \hline
$\mathcal{L}$ & $\frac{2}{3}$ & $\frac{2}{3}$ & $-\frac{2}{3}$ & $0$ & $0$ & $0$ & $-2$
& $0$ & $2$ & $2$ & $\frac1{3}$ & $-1$ & $-1$ & $1$ & $1$ & $1$\\ \hline
$A_{4}$ & $\mathbf{1}^{\prime\prime}$ & $\mathbf{1}^{\prime}$ & $\mathbf{1}$
& $\mathbf{1}$ & $\mathbf{1}$ & $\mathbf{1}$ & $\mathbf{1}$ & $\mathbf{3}$ & 
$\mathbf{1}^{\prime\prime}$ & $\mathbf{1}^{\prime}$ & $\mathbf{3}$ & $%
\mathbf{3}$ & $\mathbf{3}$ & $\mathbf{1}$ & $\mathbf{1}$ & $\mathbf{1}$ \\ 
\hline
$Z_{4}$ & $0$ & $0$ & $0$ & $0$ & $0$ & $0$ & $0$ & $0$ & $0$ & $0$ & $0$ & $%
0$ & $-1$ & $0$ & $0$ & $0$ \\ \hline
$Z_{6}$ & $0$ & $0$ & $0$ & $4$ & $4$ & $4$ & $0$ & $2$ & $0$ & $0$ & $3$ & $%
3$ & $0$ & $3$ & $3$ & $3$ \\ \hline
$Z_{16}$ & $0$ & $2$ & $4$ & $8$ & $6$ & $4$ & $4$ & $0$ & $0$ & $2$ & $0$ & 
$0$ & $0$ & $5$ & $0$ & $0$ \\ \hline
$Z_{16}^{\prime}$ & $0$ & $0$ & $0$ & $8$ & $4$ & $0$ & $0$ & $1$ & $0$ & $0$
& $0$ & $0$ & $0$ & $3$ & $8$ & $3$ \\ \hline
\end{tabular}%
\caption{Fermion assignments under $U(1)_{\mathcal{L}}\times A_{4}\times Z_{4}\times Z_{6}\times
Z_{16}\times Z_{16}^{\prime }$.}
\label{ta:fermions}
\end{table}

With the above particle content, the relevant Yukawa terms for the quark and
lepton sector invariant under the group $\mathcal{G}$, respectively, are: 
\begin{eqnarray}
-\mathcal{L}_{Y}^{\left( q\right) } &=&y^{\left( T\right) }\overline{Q}%
_{3L}\chi T_{R}+y_{33}^{\left( U\right) }\overline{Q}_{3L}\eta
U_{3R}+y_{23}^{\left( U\right) }\overline{Q}_{2L}\rho ^{\ast }U_{3R}\frac{%
\sigma ^{2}}{\Lambda ^{2}}+y_{13}^{\left( U\right) }\overline{Q}_{1L}\rho
^{\ast }U_{3R}\frac{\sigma ^{4}}{\Lambda ^{4}}  \notag \\
&&+y_{22}^{\left( U\right) }\overline{Q}_{2L}\rho ^{\ast }U_{2R}\frac{\tau
_{1}^{4}}{\Lambda ^{4}}+y_{12}^{\left( U\right) }\overline{Q}_{1L}\rho
^{\ast }U_{2R}\frac{\tau _{1}^{4}\tau _{2}}{\Lambda ^{5}}+y_{11}^{\left(
U\right) }\overline{Q}_{1L}\rho ^{\ast }U_{1R}\frac{\tau _{1}^{8}}{\Lambda
^{8}}  \notag \\
&&+y_{1}^{\left( J\right) }\overline{Q}_{1L}\chi ^{\ast
}J_{1R}+y_{2}^{\left( J\right) }\overline{Q}_{2L}\chi ^{\ast
}J_{2R}+y_{1}^{\left( D\right) }\overline{Q}_{1L}\eta ^{\ast }\left( \xi
D_{R}\right) _{\mathbf{1}^{\prime \prime }}\frac{\phi ^{6}}{\Lambda ^{7}}%
+y_{2}^{\left( D\right) }\overline{Q}_{2L}\eta ^{\ast }\left( \xi
D_{R}\right) _{\mathbf{1^{\prime }}}\frac{\phi ^{4}}{\Lambda ^{5}}  \notag \\
&&+y_{3}^{\left( D\right) }\overline{Q}_{3L}\rho \left( \xi D_{R}\right) _{%
\mathbf{1}}\frac{\phi ^{2}}{\Lambda ^{3}}+H.c,  \label{Lyq}
\end{eqnarray}%
\begin{eqnarray}
-\mathcal{L}_{Y}^{\left( l\right) } &=&y_{11}^{\left( L\right) }\left( 
\overline{L}_{L}\rho \Phi \right) _{\mathbf{\mathbf{1}}}e_{1R}\frac{\phi ^{5}%
}{\Lambda ^{6}}+y_{31}^{\left( L\right) }\left( \overline{L}_{L}\rho \Delta
\right) _{\mathbf{\mathbf{1}}}e_{1R}\frac{\phi ^{5}}{\Lambda ^{6}}%
+y_{22}^{\left( L\right) }\left( \overline{L}_{L}\rho \Xi \right) _{\mathbf{%
\mathbf{1}}}e_{2R}\frac{1}{\Lambda }.  \notag \\
&&+y_{13}^{\left( L\right) }\left( \overline{L}_{L}\rho \Phi \right) _{%
\mathbf{\mathbf{1}}}e_{3R}\frac{1}{\Lambda }+y_{33}^{\left( L\right) }\left( 
\overline{L}_{L}\rho \Delta \right) _{\mathbf{\mathbf{1}}}e_{3R}\frac{1}{%
\Lambda }+y_{\rho }\varepsilon _{abc}\left( \overline{L}_{L}^{a}\left(
L_{L}^{C}\right) ^{b}\right) _{\mathbf{3a}}\left( \rho ^{\ast }\right) ^{c}%
\frac{\zeta }{\Lambda }  \notag \\
&&+y_{\chi }^{\left( L\right) }\left( \overline{L}_{L}\chi N_{R}\right) _{%
\mathbf{\mathbf{1}}}+y_{\varphi }^{\left( N\right) }\left( N_{R}\overline{%
\Omega _{R}^{C}}\right) _{\mathbf{\mathbf{1}}}\varphi +y_{\varrho }^{\left(
\Omega \right) }\left( \Omega _{R}\overline{\Omega _{R}^{C}}\right) _{%
\mathbf{\mathbf{1}}}\varrho +y_{\Theta }^{\left( \Omega \right) }\left(
\Omega _{R}\overline{\Omega _{R}^{C}}\right) _{\mathbf{3s}}\Theta +H.c
\label{Lyl}
\end{eqnarray}%
where the dimensionless couplings in Eq. (\ref{Lyq}) and (\ref{Lyl}) are $%
\mathcal{O}(1)$ parameters. Furthermore, as it will shown in Sect. \ref%
{quarksector}, the quark assignments under the different group factors of
our model will give rise to SM quark mass textures where the CKM quark
mixing angles only arise from the up type quark sector. As indicated by the
current low energy quark flavor data encoded in the Standard parametrization
of the quark mixing matrix, the complex phase responsible for CP violation
in the quark sector is associated with the quark mixing angle in the $1$-$3$
plane. Consequently, in order to reproduce the experimental values of quark
mixing angles and CP violating phase, $y_{13}^{\left( U\right) }$ is
required to be complex. Besides that, as it will shown in Sect. \ref%
{leptonsector}, the light active neutrino sector will generate the
tribimaximal mixing matrix, whereas the charged lepton sector will give rise
to the reactor mixing angle. In order to account for CP violation in
neutrino oscillations, we will also assume that the $y_{13}^{\left( L\right)
} $ parameter is complex.

Although the flavor discrete groups in Eq. (\ref{Group}) look rather
sophisticated, each discrete group factor is crucial for generating highly
predictive SM fermion mass matrices consistent with low energy fermion
flavor data. As it will shown in Sect. \ref{leptonsector}, the predictive
textures for the lepton sectors will give rise to the experimentally
observed deviation of the tribimaximal mixing pattern. Besides that, the
resulting SM quark mass matrices will give rise to quark mixing only
emerging from the up type quark sector. This is a consequence of the $A_{4}$
flavor symmetry, which needs to be supplemented by the $A_{4}\times
Z_{4}\times Z_{6}\times Z_{16}\times Z_{16}^{\prime }$ discrete group. As we
will see in the next sections, this predictive setup can successfully
account for SM fermion masses and mixings. The inclusion of the $A_{4}$
discrete group reduces the number of parameters in the Yukawa and scalar
sector of the $SU(3)_{C}\times SU(3)_{L}\times U(1)_{X}$ model making it
more predictive. We choose $A_{4}$ since it is the smallest discrete group
with a three-dimensional irreducible representation and 3 distinct
one-dimensional irreducible representations, which allows to naturally
accommodate the three fermion families. In what follows we provide an
explanation of the role of each discrete cyclic group factor introduced in
our model. The $Z_{4}$ symmetry is the smallest cyclic symmetry that
guarantees that the renormalizable Yukawa terms for the right handed
Majorana neutrinos $\Omega _{iR}$ ($i=1,2,3$) only involve the scalar fields 
$\varrho $ and$\Theta _{i}$ ($i=1,2,3$) assumed to be real, whose VEVs are
taken to satisfy $v_{\Theta }\ll v_{\varrho }\sim \mathcal{O}(1)$ TeV. In
addition, the $Z_4$ symmetry avoids 5 dimensional Yukawa interactions of the
right handed Majorana neutrinos $\Omega _{iR}$ ($i=1,2,3$) with the scalar
fields $\xi_i$ ($i=1,2,3$) (which acquire VEVs at very high energy scale),
that could push the masses for these right handed Majorana neutrinos at very
high scale. Consequently, the $Z_4$ symmetry is crucial to have TeV scale
inverse seesaw mediators $\Omega _{iR}$ ($i=1,2,3$), which allows the
implementation of a one loop level inverse seesaw mechanism to generate
light active neutrino masses, thus giving rise to exotic pseudo-Dirac
neutrinos within the LHC reach. The $Z_{6}$ symmetry has the following
roles: 1) To separate the $A_{4}$ scalar triplet $\zeta $ participating in
the Dirac neutrino Yukawa interactions from the remaining $A_{4}$ scalar
triplets. 2) To forbid mixings between SM quarks and exotic quarks, thus
resulting in a reduction of quark sector model parameters. 3) To allow the
implementation of the one loop level inverse seesaw mechanism for the
generation of the light active neutrino masses, due to to the fact that the $%
Z_{6}$ discrete group is broken down to the preserved $Z_{2}$ symmetry. Let
us note that we use the $Z_{6}$ discrete group since it is the smallest
cyclic group that contains both the $Z_{3}$ and $Z_{2}$ symmetries. The $%
Z_{3}$ symmetry contained in $Z_{6}$ allows to decouple the exotic quarks
from the SM quarks, whereas the preserved $Z_{2}$ symmetry is crucial for
the implementation of the one loop level inverse seesaw mechanism for the
generation of the light active neutrino masses. In what concerns, the $%
Z_{16} $ symmetry, it is worth mentioning that it is crucial to generate the
observed charged fermion mass and quark mixing pattern. Let us note, that
the properties of the $Z_{N}$ groups imply that the $Z_{16}$ symmetry is the
smallest cyclic symmetry from which the Yukawa term $\overline{Q}%
_{L}^{1}\rho ^{\ast }U_{R}^{1}\frac{\tau _{1}^{8}}{\Lambda ^{8}}$ of
dimension twelve can be built, from a $\frac{\tau _{1}^{8}}{\Lambda ^{8}}$
insertion on the $\overline{Q}_{L}^{1}\rho ^{\ast }U_{R}^{1}$ operator,
crucial to get the required $\lambda ^{8}$ suppression (where $\lambda
=0.225 $ is one of the Wolfenstein parameters) needed to naturally explain
the smallness of the up quark mass, which is $\lambda ^{8}\frac{v}{\sqrt{2}}$
($\lambda =0.225$ is one of the Wolfenstein parameters) times a $\mathcal{O}%
(1) $ parameter. Furthermore, the $Z_{16}$ discrete symmetry separates the $%
A_{4} $ scalar triplet $\xi $ participating in the SM down type quark Yukawa
interactions from the remaining $A_{4}$ scalar triplets. The $Z_{16}^{\prime
}$ symmetry has the functions: 1) To select the allowed entries of the SM
quark mass matrices, thus yielding a very predictive quark sector. It is
worth mentioning that the $Z_{16}^{\prime }$ is the smallest cyclic symmetry
that allows us to get vanishing $\left( 2,1\right) $, $\left( 3,1\right) $
and $\left( 3,2\right) $ entries in the SM up type quark mass matrix. 
2) To distinguish the $A_{4}$ scalar triplet $\xi $ participating in the
quark Yukawa interactions, from the ones, i.e., $\zeta $ and $\Theta $\ that
appear in the neutrino Yukawa terms and from the $A_{4}$ scalar triplets,
i.e., $\Phi $, $\Delta $ and $\Xi $ , contributing to the charged lepton
masses, thus allowing to treat, the SM down type quark, the charged lepton
and neutrino sectors independently. 3) To separate the $A_{4}$ scalar
triplets $\Phi $ and $\Delta $ contributing to the electron and tau lepton
masses as well as to the reactor mixing angle from the $A_{4}$ scalar
triplet $\Xi $ that give rises to the muon lepton mass. This is crucial to
generate the experimentally observed deviation from the tribimaximal mixing
pattern, which in our model arises from the charged lepton sector.

Furthermore, since the breaking of the $A_{4}\times Z_{6}\times Z_{16}\times
Z_{16}^{\prime }$ discrete group gives rise to the charged fermion mass and
quark mixing pattern, we set the VEVs of the $SU(3)_{L}$ singlet scalar
fields (excepting $\varphi $ which has a vanishing vacuum expectation value)
with respect to the Wolfenstein parameter $\lambda =0.225$ and the model
cutoff $\Lambda $, as follows: 
\begin{equation}
v_{\Theta }\ll v_{\varrho }\sim v_{\zeta }\ll v_{\Xi }=\lambda ^{5}\Lambda
<v_{\Phi }=\lambda ^{4}\Lambda <v_{\Delta }=\lambda ^{3}\Lambda \ll v_{\xi
}\sim v_{\sigma }\sim v_{\phi }\sim v_{\tau _{1}}\sim v_{\tau _{2}}\sim
\lambda \Lambda  \label{VEVsinglets}
\end{equation}%
Let us note that we have assumed a hierarchy between the vacuum expectation
values of the $A_{4}$ scalar triplets, in order to simplify our analysis of
the scalar potential for the $A_{4}$ scalar triplets. That hierarchy in
their VEVs will allow us to neglect the mixings between these fields as
follows from the method of recursive expansion of Ref. \cite{Grimus:2000vj}
and to treat their scalar potentials independently. Furthermore, let us note
that we have assumed the relation $v_{\Phi }\sim \lambda v_{\Delta }$ for
the vacuum expectation values of the $A_{4}$ scalar triplets $\Phi $ and $%
\Delta $ contributing to the electron and tau lepton masses as well as to
the reactor mixing angle $\theta _{13}$. That assumption is made in order to
connect the reactor mixing parameter $\sin ^{2}\theta _{13}$ with the
Wolfenstein parameter $\lambda =0.225$, through the relation $\sin \theta
_{13}\sim \lambda $, which is suggested by the neutrino oscillation
experimental data.

In the following we comment on the possible VEV patterns for the $A_{4}$
scalar triplets $\xi $, $\zeta $, $\Phi $, $\Delta $, $\Xi $, $\Theta $.
Since the VEVs of the $A_{4}$ scalar triplets satisfy the following
hierarchy: $v_{\Theta }\ll v_{\varrho }\sim v_{\zeta }\ll v_{\Xi }<v_{\Phi
}<v_{\Delta }\ll v_{\xi }$ the mixing angles between $\xi $,\ $\Delta $, $%
\Phi $, $\Xi $, $\zeta $ and $\Theta $ are very small since they are
suppressed by the ratios of their VEVs, which is a consequence of the method
of recursive expansion proposed in Ref. \cite{Grimus:2000vj}. Thus, the
scalar potentials for the $A_{4}$ scalar triplets $\xi $, $\zeta $, $\Phi $, 
$\Delta $, $\Xi $, $\Theta $ can be treated independently. As shown in
detail in Appendix \ref{B1}, the following VEV patterns for the $A_{4}$
scalar triplets are consistent with the scalar potential minimization
equations for a large region of parameter space:

\begin{eqnarray}
\left\langle \xi \right\rangle &=&\frac{v_{\xi }}{\sqrt{3}}\left(
1,1,1\right) ,\hspace{1cm}\left\langle \Phi \right\rangle =v_{\Phi }\left(
1,0,0\right) ,\hspace{1cm}\left\langle \Delta \right\rangle =v_{\Delta
}\left( 0,0,1\right) ,  \notag \\
\left\langle \Xi \right\rangle &=&v_{\Xi }\left( 0,1,0\right) ,\hspace{1cm}%
\left\langle \zeta \right\rangle =\frac{v_{\zeta }}{\sqrt{2}}\left(
0,-1,1\right) ,\hspace{1cm}\left\langle \Theta \right\rangle =-\frac{%
v_{\Theta }}{\sqrt{5}}\left( 1,2,0\right) .  \label{VEVpattern}
\end{eqnarray}

\section{Low energy scalar potential}

\label{scalarpotential} The renormalizable low energy scalar potential of
the model under consideration is given by:%
\begin{eqnarray}
V &=&-\mu _{\chi }^{2}(\chi ^{\dagger }\chi )-\mu _{\eta }^{2}(\eta
^{\dagger }\eta )-\mu _{\rho }^{2}(\rho ^{\dagger }\rho )+\mu _{\varphi
}^{2}\varphi \varphi ^{\ast }-\mu _{\varrho }^{2}\varrho ^{2}+f_{1}\left(
\eta _{i}\chi _{j}\rho _{k}\varepsilon ^{ijk}+H.c.\right) +f_{2}\left[
\varrho \left( \varphi ^{\ast }\right) ^{2}+H.c\right]  \notag \\
&&+\lambda _{1}(\chi ^{\dagger }\chi )(\chi ^{\dagger }\chi )+\lambda
_{2}(\rho ^{\dagger }\rho )(\rho ^{\dagger }\rho )+\lambda _{3}(\eta
^{\dagger }\eta )(\eta ^{\dagger }\eta )+\lambda _{4}(\chi ^{\dagger }\chi
)(\rho ^{\dagger }\rho )+\lambda _{5}(\chi ^{\dagger }\chi )(\eta ^{\dagger
}\eta )  \notag \\
&&+\lambda _{6}(\rho ^{\dagger }\rho )(\eta ^{\dagger }\eta )+\lambda
_{7}(\chi ^{\dagger }\eta )(\eta ^{\dagger }\chi )+\lambda _{8}(\chi
^{\dagger }\rho )(\rho ^{\dagger }\chi )+\lambda _{9}(\rho ^{\dagger }\eta
)(\eta ^{\dagger }\rho )+\lambda _{10}\left( \varphi \varphi ^{\ast }\right)
^{2}  \notag \\
&&+\lambda _{11}(\eta ^{\dagger }\eta )\left( \varphi \varphi ^{\ast
}\right) +\lambda _{12}(\rho ^{\dagger }\rho )\left( \varphi \varphi ^{\ast
}\right) +\lambda _{13}(\chi ^{\dagger }\chi )\left( \varphi \varphi ^{\ast
}\right) +\lambda _{14}(\eta ^{\dagger }\eta )\varrho ^{2}+\lambda
_{15}(\rho ^{\dagger }\rho )\varrho ^{2}  \notag \\
&&+\lambda _{16}(\chi ^{\dagger }\chi )\varrho ^{2}+\lambda _{17}\varrho ^{4}
\label{V}
\end{eqnarray}
where $\varrho $ is the only real scalar. From the scalar potential given
above, one obtains that the scalar mass eigenstates are connected with the
weak scalar states by the following approximate relations 
\begin{eqnarray}
\begin{pmatrix}
G_{1}^{\pm } \\ 
H_{1}^{\pm } \\ 
\end{pmatrix}%
\simeq R_{\beta _{T}}%
\begin{pmatrix}
\rho _{1}^{\pm } \\ 
\eta _{2}^{\pm } \\ 
\end{pmatrix}
&,&\hspace{0.3cm}%
\begin{pmatrix}
G_{1}^{0} \\ 
A_{1}^{0} \\ 
\end{pmatrix}%
\simeq R_{\beta _{T}}%
\begin{pmatrix}
\zeta _{\rho } \\ 
\zeta _{\eta } \\ 
\end{pmatrix}%
,\hspace{0.3cm}%
\begin{pmatrix}
H_{1}^{0} \\ 
h^{0} \\ 
\end{pmatrix}%
\simeq R_{\alpha _{T}}%
\begin{pmatrix}
\xi _{\rho } \\ 
\xi _{\eta } \\ 
\end{pmatrix}%
,  \label{331-mass-scalar-a} \\
\begin{pmatrix}
G_{2}^{0} \\ 
H_{2}^{0} \\ 
\end{pmatrix}%
\simeq R_{1}%
\begin{pmatrix}
\chi _{1}^{0} \\ 
\eta _{3}^{0} \\ 
\end{pmatrix}
&,&\hspace{0.3cm}%
\begin{pmatrix}
G_{3}^{0} \\ 
H_{3}^{0} \\ 
H_{4}^{0}%
\end{pmatrix}%
\simeq R_{2}%
\begin{pmatrix}
\zeta _{\chi } \\ 
\xi _{\chi } \\ 
\varrho%
\end{pmatrix}%
,\hspace{0.3cm}%
\begin{pmatrix}
G_{2}^{\pm } \\ 
H_{2}^{\pm } \\ 
\end{pmatrix}%
\simeq R%
\begin{pmatrix}
\chi _{2}^{\pm } \\ 
\rho _{3}^{\pm } \\ 
\end{pmatrix}%
,\hspace{0.3cm}%
\begin{pmatrix}
A_{2}^{0} \\ 
H_{5}^{0} \\ 
\end{pmatrix}%
=I%
\begin{pmatrix}
\func{Im}\varphi \\ 
\func{Re}\varphi \\ 
\end{pmatrix}%
,  \label{331-mass-scalar-b}
\end{eqnarray}
with 
\begin{eqnarray}
R_{\alpha (\beta )} &=&\left( 
\begin{array}{cc}
\cos \alpha (\beta ) & \sin \alpha (\beta ) \\ 
-\sin \alpha (\beta ) & \cos \alpha (\beta )%
\end{array}%
\right) ,\hspace{2cm}\tan \beta =\frac{v_{\eta }}{v_{\rho }},\hspace{2cm}%
\tan 2\alpha =\frac{M_{1}^{2}}{M_{2}^{2}-M_{3}^{2}},  \notag \\
M_{1}^{2} &=&4\lambda _{6}v_{\eta }v_{\rho }+2\sqrt{2}f_{1}v_{\chi }\hspace{%
2cm}M_{2}^{2}=4\lambda _{2}v_{\rho }^{2}-\sqrt{2}f_{1}v_{\chi }\tan \beta ,%
\hspace{2cm}M_{3}^{2}=4\lambda _{3}v_{\eta }^{2}-\frac{\sqrt{2}f_{1}v_{\chi }%
}{\tan \beta }, \\
R_{1} &=&\left( 
\begin{array}{cc}
-1 & 0 \\ 
0 & 1%
\end{array}%
\right) ,\hspace{1cm}I=\left( 
\begin{array}{cc}
1 & 0 \\ 
0 & 1%
\end{array}%
\right) ,\hspace{1cm}R_{2}=\left( 
\begin{array}{ccc}
-1 & 0 & 0 \\ 
0 & \cos \gamma & \sin \gamma \\ 
0 & -\sin \gamma & \cos \gamma%
\end{array}%
\right) ,\hspace{1cm}\tan 2\gamma =\frac{4\lambda _{16}v_{\chi }v_{\varrho }%
}{4\lambda _{17}v_{\varrho }^{2}-\lambda _{1}v_{\chi }^{2}}.  \notag
\end{eqnarray}
The low energy physical scalar spectrum of our model is composed of the
following fields: 4 massive charged Higgs ($H_{1}^{\pm }$, $H_{2}^{\pm }$),
two CP-odd Higgses ($A_{1}^{0},A_{2}^{0}$), 5 neutral CP-even Higgs ($%
h^{0},H_{1}^{0},H_{3}^{0},H_{4}^{0},H_{5}^{0}$) and 2 neutral Higgs ($%
H_{2}^{0},\overline{H}_{2}^{0}$) bosons. The scalar $h^{0}$ is identified
with the SM-like $126$ GeV Higgs boson found at the LHC. It it noteworthy
that the neutral Goldstone bosons $G_{1}^{0}$, $G_{3}^{0}$, $G_{2}^{0}$ , $%
\overline{G}_{2}^{0}$ are associated to the longitudinal components of the $%
Z $, $Z^{\prime }$, $K^{0}$ and $\overline{K}^{0}$gauge bosons,
respectively. Furthermore, the charged Goldstone bosons $G_{1}^{\pm }$ and $%
G_{2}^{\pm }$ are associated to the longitudinal components of the $W^{\pm }$
and $K^{\pm } $ gauge bosons, respectively.

\section{Quark masses and mixings.}

\label{quarksector} From the quark Yukawa interactions given by Eq. (\ref%
{Lyq}) we find that the SM mass matrices for quarks take the form:

\begin{eqnarray}
M_{U} &=&\frac{v}{\sqrt{2}}\left( 
\begin{array}{ccc}
c_{1}\lambda ^{8} & b_{1}\lambda ^{5} & a_{1}\lambda ^{4} \\ 
0 & b_{2}\lambda ^{4} & a_{2}\lambda ^{2} \\ 
0 & 0 & a_{3}%
\end{array}%
\right) ,\hspace{1cm}\hspace{1cm}M_{D}=\frac{v}{\sqrt{2}}\left( 
\begin{array}{ccc}
g_{1}\lambda ^{7} & 0 & 0 \\ 
0 & g_{2}\lambda ^{5} & 0 \\ 
0 & 0 & g_{3}\lambda ^{3}%
\end{array}%
\right) R_{D},\hspace{1cm}\hspace{1cm}  \label{Mq} \\
R_{D} &=&\frac{1}{\sqrt{3}}\left( 
\begin{array}{ccc}
1 & \omega ^{2} & \omega \\ 
1 & \omega & \omega ^{2} \\ 
1 & 1 & 1%
\end{array}%
\right) ,\hspace{2cm}\omega =e^{\frac{2\pi i}{3}}.  \notag
\end{eqnarray}
where $c_{1}$, $b_{n}$ ($n=1,2$),$\ a_{i}$,$\ g_{i}$ ($i=1,2,3$) are $%
\mathcal{O}(1)$ dimensionless parameters. Here $\lambda =0.225$ is one of
the Wolfenstein parameters and $v=246$ GeV the scale of electroweak symmetry
breaking. From the SM quark mass textures given above, it follows that the
quark mixing angles only arise from the up type quark sector. Besides that,
the low energy quark flavor data indicates that the CP violating phase in
the quark sector is associated with the quark mixing angle in the 1-3 plane,
as follows from the Standard parametrization of the quark mixing matrix.
Consequently, in order to get quark mixing angles and a CP violating phase
consistent with the experimental data, we assume that all dimensionless
parameters given in Eqs. (\ref{Mq}) are real, except for $a_{1}$, taken to
be complex.

Furthermore, as follows from the different $Z_{6}$ charge assignments for
the quark fields, the exotic quarks do not mix with the SM quarks. We find
that the exotic quark masses are given by: 
\begin{equation}
m_{T}=y^{\left( T\right) }\frac{v_{\chi }}{\sqrt{2}},\hspace{1cm}%
m_{J^{1}}=y_{1}^{\left( J\right) }\frac{v_{\chi }}{\sqrt{2}}=\frac{%
y_{1}^{\left( J\right) }}{y^{\left( T\right) }}m_{T},\hspace{1cm}%
m_{J^{2}}=y_{2}^{\left( J\right) }\frac{v_{\chi }}{\sqrt{2}}=\frac{%
y_{2}^{\left( J\right) }}{y^{\left( T\right) }}m_{T}.  \label{mexotics}
\end{equation}

The obtained values for the physical quark mass spectrum \cite%
{Bora:2012tx,Xing:2007fb}, mixing angles and Jarlskog invariant \cite%
{Olive:2016xmw} are consistent with their experimental data, as shown in
Table \ref{Tab}, starting from the following benchmark point:

\begin{eqnarray}
c_{1} &\simeq &1.233,\hspace{1cm}b_{1}\simeq 1.429,\hspace{1cm}b_{2}\simeq
1.392,\hspace{1cm}\left\vert a_{1}\right\vert \simeq 3.353,\hspace{1cm}%
a_{2}\simeq -0.800  \notag \\
a_{3} &\simeq &0.999,,\hspace{1cm}\arg \left( a_{1}\right) \simeq
-156.74^{\circ },\hspace{1cm}g_{1}\simeq 0.565,\hspace{1cm}g_{2}\simeq 0.570,%
\hspace{1cm}g_{3}\simeq 1.416,
\end{eqnarray}

\begin{table}[tbh]
\begin{center}
\begin{tabular}{c|l|l}
\hline\hline
Observable & Model value & Experimental value \\ \hline
$m_{u}(MeV)$ & \quad $1.39$ & \quad $1.45_{-0.45}^{+0.56}$ \\ \hline
$m_{c}(MeV)$ & \quad $657$ & \quad $635\pm 86$ \\ \hline
$m_{t}(GeV)$ & \quad $174$ & \quad $172.1\pm 0.6\pm 0.9$ \\ \hline
$m_{d}(MeV)$ & \quad $2.9$ & \quad $2.9_{-0.4}^{+0.5}$ \\ \hline
$m_{s}(MeV)$ & \quad $57.7$ & \quad $57.7_{-15.7}^{+16.8}$ \\ \hline
$m_{b}(GeV)$ & \quad $2.82$ & \quad $2.82_{-0.04}^{+0.09}$ \\ \hline
$\sin \theta _{12}$ & \quad $0.220$ & \quad $0.2254$ \\ \hline
$\sin \theta _{23}$ & \quad $0.0414$ & \quad $0.0413$ \\ \hline
$\sin \theta _{13}$ & \quad $0.00354$ & \quad $0.00351$ \\ \hline
$\delta $ & \quad $72^{\circ }$ & \quad $68^{\circ }$ \\ \hline\hline
\end{tabular}%
\end{center}
\caption{Model and experimental values of the quark masses and CKM
parameters.}
\label{Tab}
\end{table}
In Table \ref{Tab} we show the model and experimental values for the
physical observables of the quark sector. We use the $M_{Z}$-scale
experimental values of the quark masses given by Ref. \cite{Bora:2012tx}
(which are similar to those in \cite{Xing:2007fb}). The experimental values
of the CKM parameters are taken from Ref. \cite{Olive:2016xmw}. As indicated
by Table \ref{Tab}, the obtained quark masses, quark mixing angles, and CP
violating phase are consistent with the low energy quark flavor data.

\begin{figure}[tbp]
\includegraphics[width=0.5\textwidth]{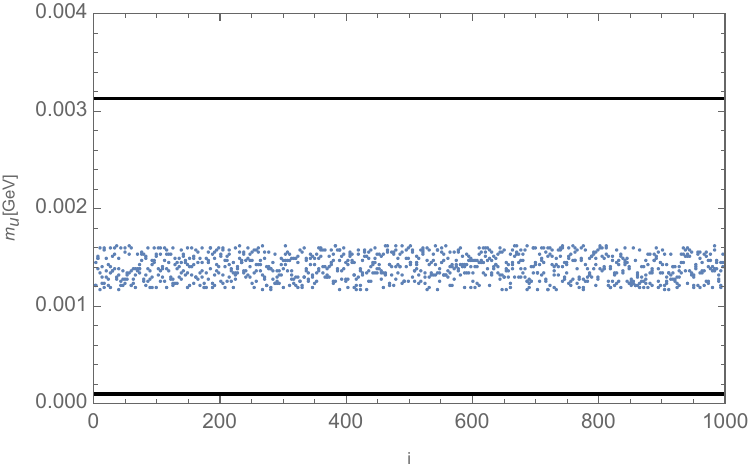}\includegraphics[width=0.5%
\textwidth]{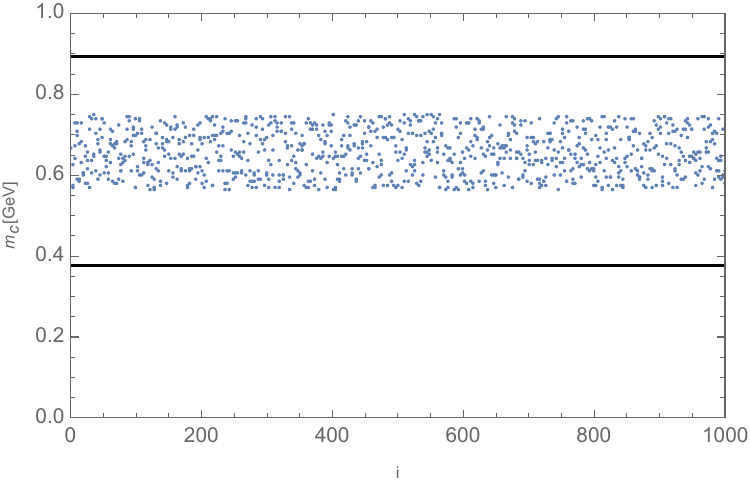}\newline
\includegraphics[width=0.5\textwidth]{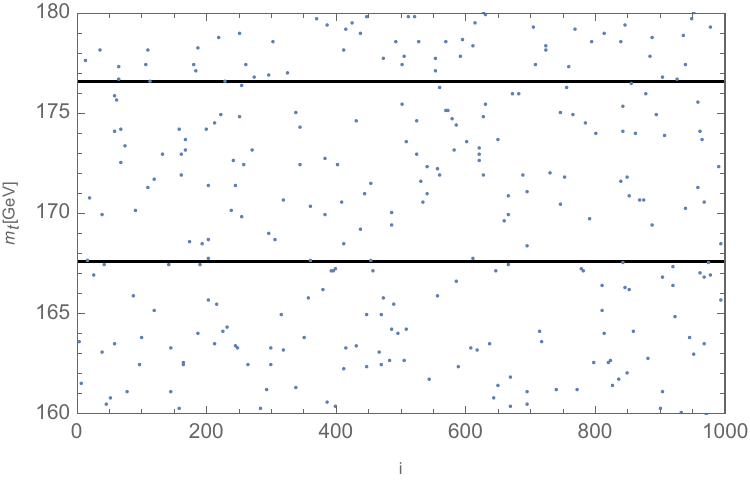}\includegraphics[width=0.5%
\textwidth]{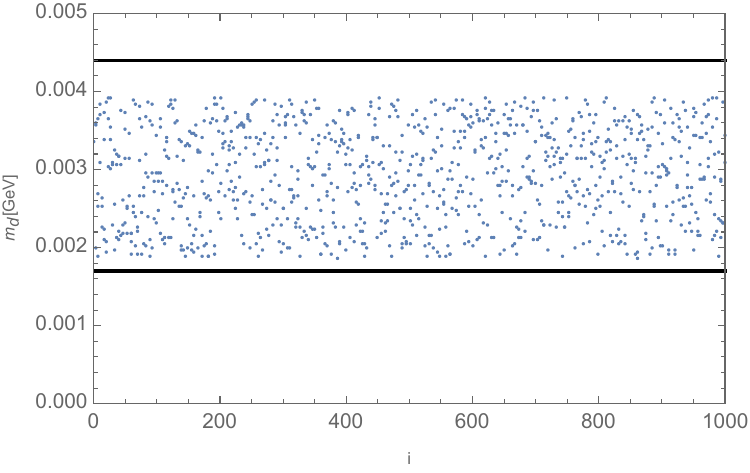}\newline
\includegraphics[width=0.5\textwidth]{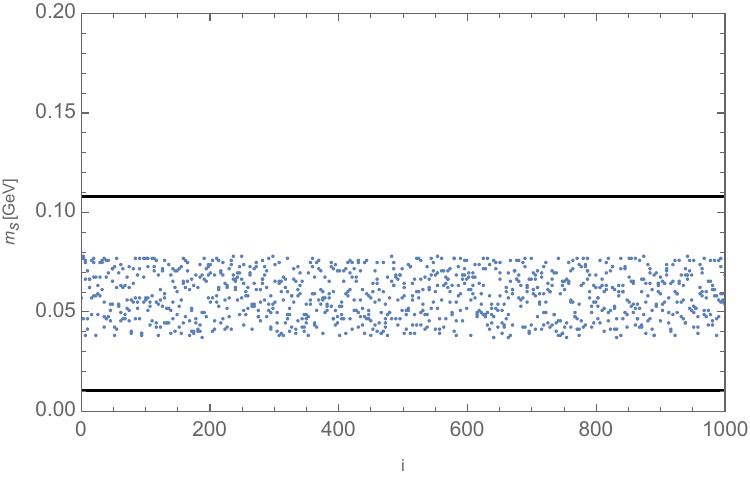}\includegraphics[width=0.5%
\textwidth]{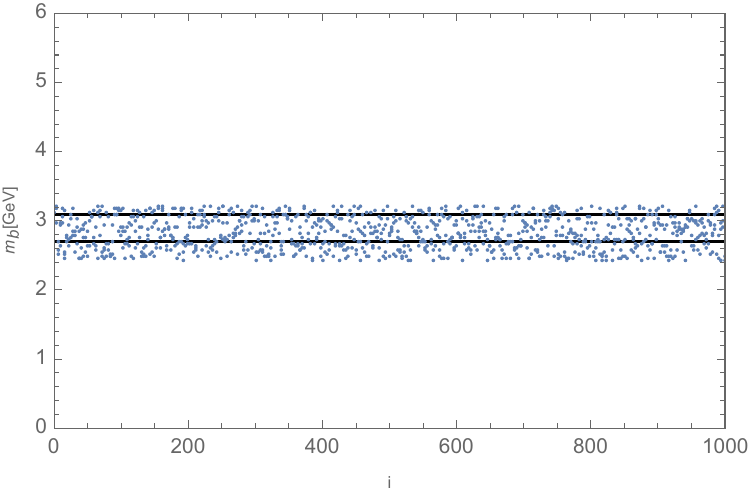}\newline
\caption{SM quark masses randomly generated. The horizonal lines are the
minimum and maximum values of the neutrino mass squared splittings inside
the $3\protect\sigma$ experimentally allowed range.}
\label{mq}
\end{figure}
\begin{figure}[tbp]
\includegraphics[width=0.5\textwidth]{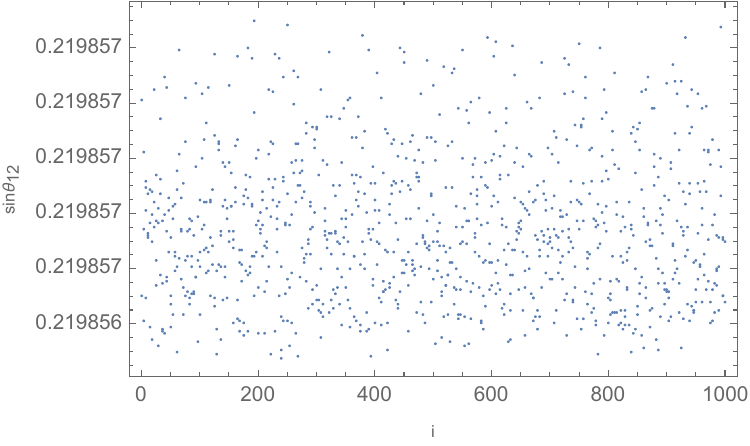}\includegraphics[width=0.5%
\textwidth]{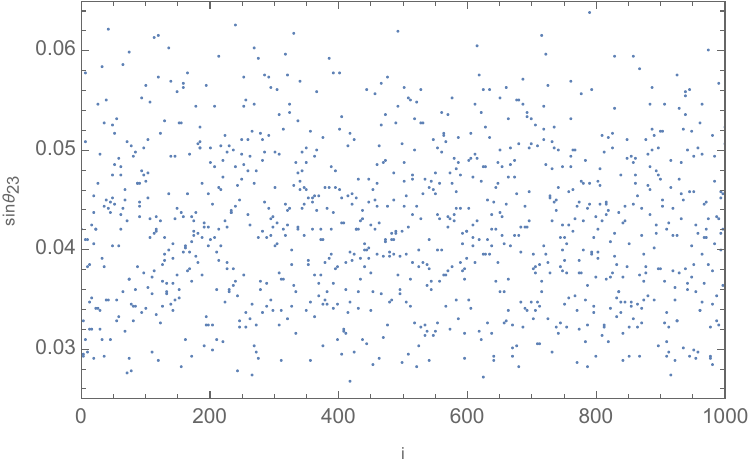}\newline
\includegraphics[width=0.5\textwidth]{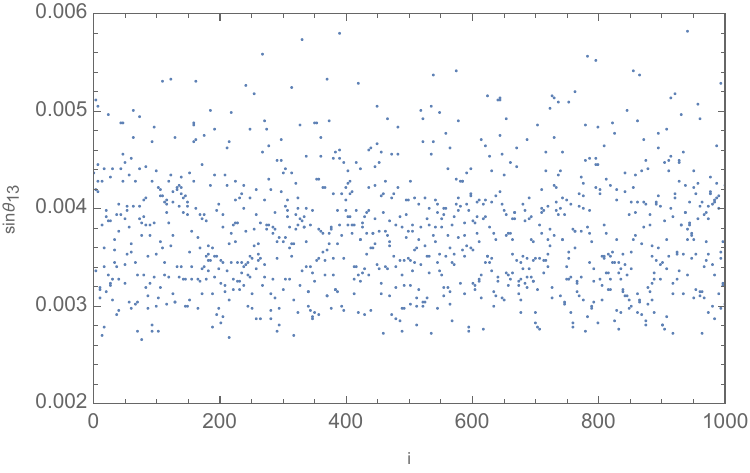}\includegraphics[width=0.5%
\textwidth]{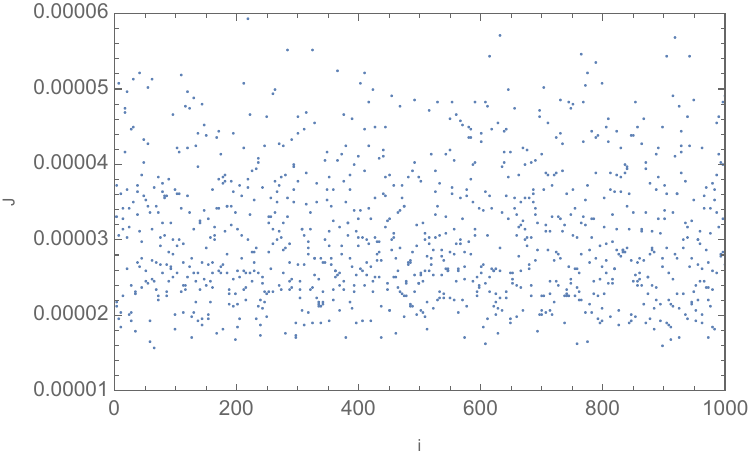}\newline
\caption{Quark mixing parameters and Jarlskog invariant randomly generated.}
\label{CKM}
\end{figure}

In order to study the sensitivity of the obtained values for the SM quark
masses, and CKM parameters under small variations around the best-fit values
(maximum variation of $+0.2$, minimum of $-0.2$), we show in Figures \ref{mq}
and \ref{CKM} the predicted SM quark masses and CKM parameters,
respectively, as functions of the iteration. We find that a slight deviation
from the best-fit values, keeps all the obtained SM quark masses, with the
exception of the top and bottom quark masses, inside the $3\sigma$
experimentally allowed range. In what regards the top and bottom quark
masses, a large amount of points are inside the $3\sigma$ experimentally
allowed range. The points outside the $3\sigma$ experimentally allowed
range, correspond to values close to the lower and upper experimental bounds
of the bottom quark mass. In what concerns the quark mixing angles and
Jarlskog invariant we find that a slight deviation from the best-fit values
keeps the these CKM parameters within the same order of magnitude.
Consequently, our model is very predictive for the quark sector.

On the other hand, from the SM quark textures, it follows that in order to
obtain realistic SM quark masses and mixing angles without requiring a
strong hierarchy among the Yukawa couplings, one should have $v_{\rho }\sim
v_{\eta }$, which implies that $\tan \beta \sim \mathcal{O}(1)$.
Furthermore, as the $h^{0}b\bar{b}$ coupling is proportional to $\frac{\sin
\alpha }{\cos \beta }$, in order to get a $h^{0}b\bar{b}$ coupling close to
the SM expectation, we have $\alpha \sim \beta \pm \frac{\pi }{2}$. In what
follows we briefly comment about the phenomenological implications of our
model in the concerning to the flavor changing processes involving quarks.
As previously mentioned, the different $Z_{6}$ charge assignments for SM and
exotic right handed quark fields imply the absence of mixing between them.
The absence of mixings between the SM and exotic quarks will imply that the
exotic fermions will not exhibit flavor changing decays into SM quarks and
gauge (or Higgs) bosons. After being pair produced they will decay into the
SM quarks and the intermediate states of heavy gauge bosons, which in turn
decay into the pairs of the SM fermions, see e.g. \cite{Cabarcas:2008ys}.
The precise signature of the decays of the exotic quarks depends on details
of the spectrum and other parameters of the model. The present lower limits
on the $Z^{\prime }$ gauge boson mass in $3\text{-}3\text{-}1$ models
arising from LHC searches, reach around $2.5$ \text{TeV} \cite%
{Salazar:2015gxa}. These bounds can be translated into limits of about 6.3
TeV on the $SU(3)_{C}\otimes SU\left( 3\right) _{L}\otimes U\left( 1\right)
_{X}$ gauge symmetry breaking scale $v_{\chi }$. Furthermore, electroweak
data from the decays $B_{s,d}\rightarrow \mu ^{+}\mu ^{-}$ and $%
B_{d}\rightarrow K^{\ast }(K)\mu ^{+}\mu ^{-}$ set lower bounds on the $%
Z^{\prime }$ gauge boson mass ranging from $1$ TeV up to $3$ TeV \cite%
{CarcamoHernandez:2005ka,Martinez:2008jj,Buras:2013dea,Buras:2014yna,Buras:2012dp}%
. The exotic quarks can be pair produced at the LHC via Drell-Yan and gluon
fusion processes mediated by charged gauge bosons and gluons, respectively.
A detailed study of the exotic quark production at the LHC and the exotic
quark decay modes is beyond the scope of this work and is deferred for a
future publication.

\section{Lepton masses and mixings.}

\label{leptonsector}

From Eqs. (\ref{Lyl}), (\ref{VEVsinglets}), (\ref{VEVpattern}) and using the
product rules of the $A_{4}$ group given in Appendix \ref{A4}, we find that
the charged lepton mass matrix is given by: 
\begin{equation}
M_{l}=\frac{v}{\sqrt{2}}\left( 
\begin{array}{ccc}
x_{1}\lambda ^{9} & 0 & z_{1}\lambda ^{4} \\ 
0 & y\lambda ^{5} & 0 \\ 
x_{2}\lambda ^{8} & 0 & z_{2}\lambda ^{3}%
\end{array}%
\right),\label{Ml}
\end{equation}%
where $x_{n}$, $y$, $z_{n}$ ($n=1,2$)\ are $\mathcal{O}(1)$ dimensionless
parameters, assumed to be real, excepting $z_{1}$, taken to be complex, in
order to generate a nonvanishing leptonic Dirac CP violating phase.
Specifically, for the sake of simplicity we take $z_{1}$ as $%
z_{1}=\left\vert z_{1}\right\vert e^{i\kappa }$.

The matrix $M_{l}M_{l}^{\dagger }$ is diagonalized by a rotation matrix $%
R_{l}$ according to:%
\begin{equation}
R_{l}^{\dagger }M_{l}M_{l}^{\dagger }R_{l}=\left( 
\begin{array}{ccc}
m_{e}^{2} & 0 & 0 \\ 
0 & m_{\mu }^{2} & 0 \\ 
0 & 0 & m_{\tau }^{2}%
\end{array}%
\right) ,\hspace{1cm}\hspace{1cm}R_{l}=\left( 
\begin{array}{ccc}
\cos \theta _{l} & 0 & -e^{i\kappa }\sin \theta _{l} \\ 
0 & 1 & 0 \\ 
e^{-i\kappa }\sin \theta _{l} & 0 & \cos \theta _{l}%
\end{array}%
\right) \allowbreak ,\hspace{1cm}\hspace{1cm}\tan \theta _{l}\simeq -\frac{%
\left\vert z_{1}\right\vert }{z_{2}}\lambda ,  \label{Rl}
\end{equation}%
where the charged lepton masses are approximately given by: 
\begin{equation}
m_{e}\simeq \sqrt{\frac{x_{2}^{2}\allowbreak \left\vert z_{1}\right\vert
^{2}+x_{1}^{2}z_{2}^{2}-2x_{1}x_{2}z_{1}z_{2}\cos \kappa }{z_{2}^{2}+\lambda
^{2}\left\vert z_{1}\right\vert ^{2}}}\lambda ^{9}\frac{v}{\sqrt{2}}%
\allowbreak ,\hspace{1cm}m_{\mu }=y\lambda ^{5}\frac{v}{\sqrt{2}},\hspace{1cm%
}m_{\tau }\simeq \sqrt{z_{2}^{2}+z_{1}^{2}\lambda ^{2}}\lambda ^{3}\frac{v}{%
\sqrt{2}}.
\end{equation}

It is worth mentioning that the charged lepton masses are connected with the
electroweak symmetry breaking scale $v=246$ GeV by their scalings with
powers of the Wolfenstein parameter $\lambda =0.225$, with $\mathcal{O}(1)$
coefficients. This is consistent with our previous assumption made in Eq. (%
\ref{VEVsinglets}) regarding the size of the VEVs for the $SU(3)_{L}$
singlet scalars appearing in the charged fermion Yukawa terms. Furthermore,
it is noteworthy that the mixing angle $\theta _{l}$ in the charged lepton
sector is large, which gives rise to an important contribution to the
leptonic mixing matrix, coming from the mixing of charged leptons.

\begin{figure}[t]
\center
\vspace{0.8cm} \subfigure{\includegraphics[width=0.9%
\textwidth]{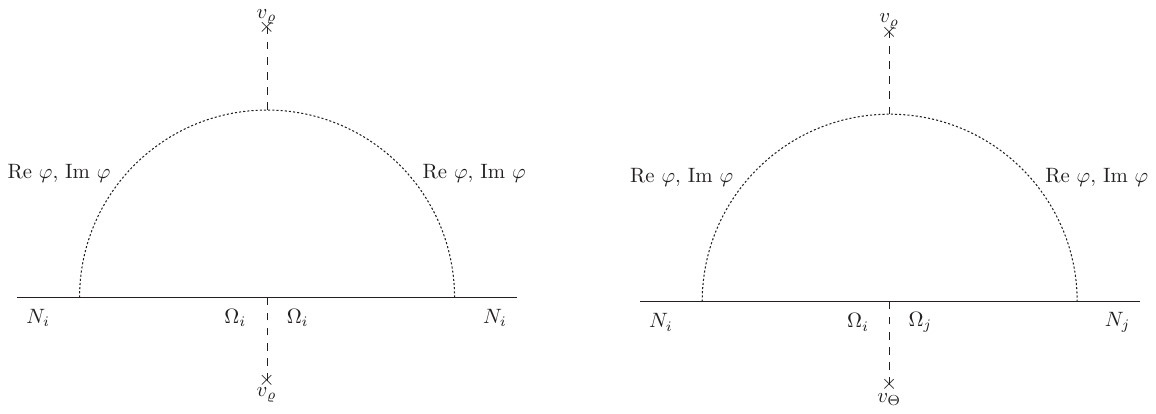}}\vspace{-15cm}
\caption{Loop Feynman diagrams contributing to the entries of the Majorana
neutrino mass submatrix $M_R$. Here $i,j=1,2,3$ and $i\ne j$. }
\label{LoopMR}
\end{figure}

Regarding the neutrino sector, from the Eq. (\ref{Lyl}), we find the
following neutrino mass terms: 
\begin{equation}
-\mathcal{L}_{mass}^{\left( \nu \right) }=\frac{1}{2}\left( 
\begin{array}{ccc}
\overline{\nu _{L}^{C}} & \overline{\nu _{R}} & \overline{N_{R}}%
\end{array}%
\right) M_{\nu }\left( 
\begin{array}{c}
\nu _{L} \\ 
\nu _{R}^{C} \\ 
N_{R}^{C}%
\end{array}%
\right) +H.c,  \label{Lnu}
\end{equation}%
where the $A_{4}$ family symmetry constrains the neutrino mass matrix to be
of the form: 
\begin{equation}
M_{\nu }=\left( 
\begin{array}{ccc}
0_{3\times 3} & M_{\nu D} & 0_{3\times 3} \\ 
M_{\nu D}^{T} & 0_{3\times 3} & M_{\chi } \\ 
0_{3\times 3} & M_{\chi }^{T} & M_{R}%
\end{array}%
\right)  \label{Mnu}
\end{equation}%
where the submatrices $M_{\nu D}$ and $M_{\chi }$ are generated at tree
level from the nonrenormalizable $\varepsilon _{abc}\left( \overline{L}%
_{L}^{a}\left( L_{L}^{C}\right) ^{b}\right) _{\mathbf{3a}}\left( \rho ^{\ast
}\right) ^{c}\frac{\zeta }{\Lambda }$ and renormalizable $\left( \overline{L}%
_{L}\chi N_{R}\right) _{\mathbf{\mathbf{1}}}$ Yukawa terms, respectively,
whereas the submatrix $M_{R}$ arises from a one loop level radiative seesaw
mechanism mediated by the massive right handed Majorana neutrinos $\Omega
_{i}$ ($i=1,2,3$) and the real $\func{Re}\varphi $ and imaginary $\func{Im}%
\varphi $ parts of the $Z_{6}$ charged scalar field $\varphi $. As
previously mentioned, the facts that the $Z_{6}$ discrete group is broken
down to the preserved $Z_{2}$ symmetry and the $SU(3)_{L}$ singlet scalar
field $\varphi $ (which appears in the neutrino Yukawa interaction $\left( 
\overline{N}_{R}\Omega \right) _{\mathbf{\mathbf{1}}}\varphi $) has a $Z_{6}$
charge corresponding to a nontrivial $Z_{2}$ charge, implies that this
scalar does not acquire a vacuum expectation value, thus generating the
submatrix $M_{R}$ only at one loop level. The one loop Feynman diagrams
contributing to the entries of the Majorana neutrino mass submatrix $M_R$
are shown in Figure \ref{LoopMR}. The submatrices $M_{\nu D}$, $M_{\chi }$
and $M_{R}$ are given by: 
\begin{eqnarray}
M_{\nu D} &=&\frac{y_{\rho }v_{\zeta }v_{\rho }}{\sqrt{2}\Lambda }\left( 
\begin{array}{ccc}
0 & 1 & 1 \\ 
-1 & 0 & 0 \\ 
-1 & 0 & 0%
\end{array}%
\right) ,\hspace{1cm}M_{\chi }=y_{\chi }^{\left( L\right) }\frac{v_{\chi }}{%
\sqrt{2}}\left( 
\begin{array}{ccc}
1 & 0 & 0 \\ 
0 & 1 & 0 \\ 
0 & 0 & 1%
\end{array}%
\right) , \\
M_{R} &=&\left( 
\begin{array}{ccc}
y_{\varrho }^{\left( \Omega \right) }F\left( y_{\varrho }^{\left( \Omega
\right) }v_{\varrho },m_{R},m_{I}\right) v_{\varrho } & 0 & -2y_{\Theta
}^{\left( \Omega \right) }F\left( -2y_{\Theta }^{\left( \Omega \right) }%
\frac{v_{\Theta }}{\sqrt{5}},m_{R},m_{I}\right) \frac{v_{\Theta }}{\sqrt{5}}
\\ 
0 & y_{\varrho }^{\left( \Omega \right) }F\left( y_{\varrho }^{\left( \Omega
\right) }v_{\varrho },m_{R},m_{I}\right) v_{\varrho } & -y_{\Theta }^{\left(
\Omega \right) }F\left( -y_{\Theta }^{\left( \Omega \right) }\frac{v_{\Theta
}}{\sqrt{5}},m_{R},m_{I}\right) \frac{v_{\Theta }}{\sqrt{5}} \\ 
-2y_{\Theta }^{\left( \Omega \right) }F\left( -2y_{\Theta }^{\left( \Omega
\right) }\frac{v_{\Theta }}{\sqrt{5}},m_{R},m_{I}\right) \frac{v_{\Theta }}{%
\sqrt{5}} & -y_{\Theta }^{\left( \Omega \right) }F\left( -y_{\Theta
}^{\left( \Omega \right) }\frac{v_{\Theta }}{\sqrt{5}},m_{R},m_{I}\right) 
\frac{v_{\Theta }}{\sqrt{5}} & y_{\varrho }^{\left( \Omega \right) }F\left(
y_{\varrho }^{\left( \Omega \right) }v_{\varrho },m_{R},m_{I}\right)
v_{\varrho }%
\end{array}%
\right) ,  \notag
\end{eqnarray}

where:%
\begin{eqnarray}
m_{R} &=&m_{\func{Re}\varphi }=\sqrt{\mu _{\varphi }^{2}+\frac{1}{2}\left(
\lambda _{11}v_{\eta }^{2}+\lambda _{12}v_{\rho }^{2}+\lambda _{13}v_{\chi
}^{2}\right) +2f_{2}v_{\varrho }}{\small ,}  \notag \\
m_{I} &=&m_{\func{Im}\varphi }=\sqrt{\mu _{\varphi }^{2}+\frac{1}{2}\left(
\lambda _{11}v_{\eta }^{2}+\lambda _{12}v_{\rho }^{2}+\lambda _{13}v_{\chi
}^{2}\right) -2f_{2}v_{\varrho }}{\small ,}  \label{mphi}
\end{eqnarray}

and the following function has been introduced \cite{Ma:2006km}: 
\begin{equation}
F\left( m_{1},m_{2},m_{3}\right) =\frac{\left( y_{\varphi }^{\left( N\right)
}\right) ^{2}}{16\pi ^{2}}\left[ \frac{m_{2}^{2}}{m_{2}^{2}-m_{1}^{2}}\ln
\left( \frac{m_{2}^{2}}{m_{1}^{2}}\right) -\frac{m_{3}^{2}}{%
m_{3}^{2}-m_{1}^{2}}\ln \left( \frac{m_{3}^{2}}{m_{1}^{2}}\right) \right] 
{\small ,}
\end{equation}

In order to connect the neutrino mass squared splittings with the quark
mixing parameters and motivated by the relation $\Delta m_{13}^{2}\sim
\lambda ^{4}\mathcal{O}(1)eV^{2}$, we set $v_{\Theta }$ $\sim \lambda
^{4}v_{\varrho }$. In addition, for the sake of simplicity we assume that
the $Z_{6}$ charged $SU(3)_{L}$ singlet scalar field $\varphi $ is heavier
than the right handed Majorana neutrinos $\Omega _{i}$ ($i=1,2,3$), in such
a way that we can restrict to the scenario:

\begin{equation}
m_{\func{Re}\varphi }^{2},m_{\func{Im}\varphi }^{2}\gg \left(y_{\varrho }^{\left(
\Omega \right)}\right)^{2}v_{\varrho }^{2}\gg
v_{\Theta }^{2}\sim \lambda ^{8}v_{\varrho }^{2}
\end{equation}%
for which the submatrix $M_{R}$ takes the form:

\begin{eqnarray}
M_{R} &\simeq &\frac{\left( y_{\varphi }^{\left( N\right) }\right)
^{2}\left( m_{\func{Re}\varphi }^{2}-m_{\func{Im}\varphi }^{2}\right) }{8\pi
^{2}\left( m_{\func{Re}\varphi }^{2}+m_{\func{Im}\varphi }^{2}\right) }%
\left( 
\begin{array}{ccc}
y_{\varrho }^{\left( \Omega \right) }v_{\varrho } & 0 & -2y_{\Theta
}^{\left( \Omega \right) }\frac{v_{\Theta }}{\sqrt{5}} \\ 
0 & y_{\varrho }^{\left( \Omega \right) }v_{\varrho } & -y_{\Theta }^{\left(
\Omega \right) }\frac{v_{\Theta }}{\sqrt{5}} \\ 
-2y_{\Theta }^{\left( \Omega \right) }\frac{v_{\Theta }}{\sqrt{5}} & 
-y_{\Theta }^{\left( \Omega \right) }\frac{v_{\Theta }}{\sqrt{5}} & 
y_{\varrho }^{\left( \Omega \right) }v_{\varrho }%
\end{array}%
\right)  \notag \\
&=&\frac{\left( y_{\varphi }^{\left( N\right) }\right) ^{2}f_{2}v_{\varrho }%
}{4\pi ^{2}\left[ \mu _{\varphi }^{2}+\frac{1}{2}\left( \lambda _{11}v_{\eta
}^{2}+\lambda _{12}v_{\rho }^{2}+\lambda _{13}v_{\chi }^{2}\right) \right] }%
\left( 
\begin{array}{ccc}
y_{\varrho }^{\left( \Omega \right) }v_{\varrho } & 0 & -2y_{\Theta
}^{\left( \Omega \right) }\frac{v_{\Theta }}{\sqrt{5}} \\ 
0 & y_{\varrho }^{\left( \Omega \right) }v_{\varrho } & -y_{\Theta }^{\left(
\Omega \right) }\frac{v_{\Theta }}{\sqrt{5}} \\ 
-2y_{\Theta }^{\left( \Omega \right) }\frac{v_{\Theta }}{\sqrt{5}} & 
-y_{\Theta }^{\left( \Omega \right) }\frac{v_{\Theta }}{\sqrt{5}} & 
y_{\varrho }^{\left( \Omega \right) }v_{\varrho }%
\end{array}%
\right)  \notag \\
&=&\left( 
\begin{array}{ccc}
\gamma _{1} & 0 & -2\gamma _{2}\lambda ^{4} \\ 
0 & \gamma _{1} & -\gamma _{2}\lambda ^{4} \\ 
-2\gamma _{2}\lambda ^{4} & -\gamma _{2}\lambda ^{4} & \gamma _{1}%
\end{array}%
\right) m_{R}{\small ,}
\end{eqnarray}%
where $\gamma _{1}$ and $\gamma _{2}$ are $\mathcal{O}(1)$ dimensionless
parameters, assumed to be real for simplicity. Furthermore, $m_{R}$ is the
mass scale for the Majorana neutrinos $N_{i}$ ($i=1,2,3$), which sets the
scale of breaking of lepton number.

As shown in detail in Ref. \cite{Catano:2012kw}, the full rotation matrix
that diagonalizes the neutrino mass matrix $M_{\nu }$ is approximately given
by: 
\begin{equation}
\mathbb{U}=%
\begin{pmatrix}
V_{\nu } & B_{3}U_{\chi } & B_{2}U_{R} \\ 
-\frac{(B_{2}^{\dagger }+B_{3}^{\dagger })}{\sqrt{2}}V_{\nu } & \frac{(1-S)}{%
\sqrt{2}}U_{\chi } & \frac{(1+S)}{\sqrt{2}}U_{R} \\ 
-\frac{(B_{2}^{\dagger }-B_{3}^{\dagger })}{\sqrt{2}}V_{\nu } & \frac{(-1-S)%
}{\sqrt{2}}U_{\chi } & \frac{(1-S)}{\sqrt{2}}U_{R}%
\end{pmatrix}%
,  \label{U}
\end{equation}%
where 
\begin{equation}
S=-\frac{1}{2\sqrt{2}y_{\chi }^{\left( L\right) }v_{\chi }}M_{R},\hspace{1cm}%
\hspace{1cm}B_{2}\simeq B_{3}\simeq \frac{1}{y_{\chi }^{\left( L\right)
}v_{\chi }}M_{D}^{\ast },
\end{equation}%
and the physical neutrino mass matrices are: 
\begin{eqnarray}
M_{\nu }^{\left( 1\right) } &=&M_{\nu D}\left( M_{\chi }^{T}\right)
^{-1}M_{R}M_{\chi }^{-1}M_{\nu D}^{T},  \label{Mnu1} \\
M_{\nu }^{\left( 2\right) } &=&-\frac{1}{2}\left( M_{\chi }+M_{\chi
}^{T}\right) +\frac{1}{2}M_{R},\hspace{1cm}\hspace{1cm}M_{\nu }^{\left(
3\right) }=\frac{1}{2}\left( M_{\chi }+M_{\chi }^{T}\right) +\frac{1}{2}%
M_{R},  \label{Mnu2}
\end{eqnarray}%
where $M_{\nu }^{\left( 1\right) }$ is the light active neutrino mass matrix
whereas $M_{\nu }^{\left( 2\right) }$ and $M_{\nu }^{\left( 3\right) }$ are
the exotic Dirac neutrino mass matrices. It is worth mentioning that
physical neutrino spectrum consists of three light active neutrinos and six
exotic neutrinos. The exotic neutrinos are pseudo-Dirac, with masses $\sim\pm v_{\chi }\sim\mathcal{O}(1)$ TeV and a small splitting $\sim m_{R}$. This scenario is much more interesting than the one proposed in Ref. \cite{Hernandez:2016eod} where the sterile neutrinos are very much outside the LHC reach since their masses are extremelly large, thus giving rise a double seesaw mechanism for the light active neutrino masses instead of the radiative inverse seesaw mechanism proposed in this work. Furthermore, $V_{\nu }$, $%
U_{R}$ and $U_{\chi }$ are the rotation matrices which diagonalize $M_{\nu
}^{\left( 1\right) }$, $M_{\nu }^{\left( 2\right) }$ and $M_{\nu }^{\left(
3\right) }$, respectively. It is worth mentioning that the heavy quasi Dirac
neutrinos can be produced in pairs at the LHC, via a Drell-Yan mechanism
mediated by a heavy non Standard Model neutral gauge boson $Z^{\prime }$.
The heavy quasi Dirac neutrinos can decay into a Standard Model charged
lepton and a $W$ gauge boson, due to their mixings with the light active
neutrinos. Thus, the observation of an excess of events in the dilepton
final states with respect to the SM background, would be a signal supporting
this model at the LHC and can be used to distiguish this model from the one proposed in Ref. \cite{Hernandez:2016eod}. A detailed study of the collider phenomenology of this model is beyond the scope of the present paper and is left for future studies.

\quad From Eq. (\ref{Mnu1}) it follows that the light active neutrino mass
matrix is given by: 
\begin{equation}
M_{\nu }^{\left( 1\right) }=\left( 
\begin{array}{ccc}
2\left( \gamma _{1}-\gamma _{2}\lambda ^{4}\right)  & 2\gamma _{2}\lambda
^{4} & 2\gamma _{2}\lambda ^{4} \\ 
2\gamma _{2}\lambda ^{4} & \gamma _{1} & \gamma _{1} \\ 
2\gamma _{2}\lambda ^{4} & \gamma _{1} & \gamma _{1}%
\end{array}%
\right) \frac{2y_{\rho }^{2}v_{\rho }^{2}v_{\zeta }^{2}m_{R}}{\left( y_{\chi
}^{\left( L\right) }\right) ^{2}v_{\chi }^{2}\Lambda ^{2}}=\left( 
\begin{array}{ccc}
2\left( A_{1\nu }-A_{2\nu }\lambda ^{4}\right)  & 2A_{2\nu }\lambda ^{4} & 
2A_{2\nu }\lambda ^{4} \\ 
2A_{2\nu }\lambda ^{4} & A_{1\nu } & A_{1\nu } \\ 
2A_{2\nu }\lambda ^{4} & A_{1\nu } & A_{1\nu }%
\end{array}%
\right) ,  \notag
\end{equation}%
Let us note that the smallness of the active neutrino masses arises from
their scaling with inverse powers of the high energy cutoff $\Lambda $ as
well as from their linear dependence on the loop induced mass scale $m_{R}$
for the Majorana neutrinos $N_{i}$ ($i=1,2,3$). 

%
The light active neutrino mass matrix $M_{\nu }^{\left( 1\right) }$ is
diagonalized by a unitary rotation matrix $R_{\nu }$, according to:

\begin{equation}
R_{\nu }^{T}M_{\nu }^{\left( 1\right) }R_{\nu }=\left\{ 
\begin{array}{l}
\left( 
\begin{array}{ccc}
0 & 0 & 0 \\ 
0 & 2\left( A_{1\nu }-2A_{2\nu }\lambda ^{4}\right) & 0 \\ 
0 & 0 & 2\left( A_{1\nu }+A_{2\nu }\lambda ^{4}\right)%
\end{array}%
\right) ,\hspace{1cm}R_{\nu }=\left( 
\begin{array}{ccc}
0 & \frac{-2}{\sqrt{6}} & \frac{1}{\sqrt{3}} \\ 
-\frac{1}{\sqrt{2}} & \frac{1}{\sqrt{6}} & \frac{1}{\sqrt{3}} \\ 
\frac{1}{\sqrt{2}} & \frac{1}{\sqrt{6}} & \frac{1}{\sqrt{3}}%
\end{array}%
\right) ,\hspace{1cm}\mbox{for \ \ \ \ NH}\ \  \\ 
\left( 
\begin{array}{ccc}
2\left( A_{1\nu }-2A_{2\nu }\lambda ^{4}\right) & 0 & 0 \\ 
0 & 2\left( A_{1\nu }+A_{2\nu }\lambda ^{4}\right) & 0 \\ 
0 & 0 & 0%
\end{array}%
\right) ,\hspace{1cm}R_{\nu }=\left( 
\begin{array}{ccc}
\frac{-2}{\sqrt{6}} & \frac{1}{\sqrt{3}} & 0 \\ 
\frac{1}{\sqrt{6}} & \frac{1}{\sqrt{3}} & -\frac{1}{\sqrt{2}} \\ 
\frac{1}{\sqrt{6}} & \frac{1}{\sqrt{3}} & \frac{1}{\sqrt{2}}%
\end{array}%
\right) ,\hspace{1cm}\mbox{for \ \ \ \ IH}%
\end{array}%
\right.  \label{Rnu}
\end{equation}

Consequently, the light active neutrino spectrum is composed of one massless
neutrino and two active neutrinos, whose masses are determined from the
experimental values of the neutrino mass squared splittings.

From Eqs. (\ref{Rl}) and Eqs. (\ref{Rnu}), it follows that the normal
hierarchy scenario leads to a too value for the large reactor mixing angle,
which is disfavored by the neutrino oscillation experimental data. Thus, the
normal neutrino mass hierarchy scenario of our model is ruled out by the
current data on neutrino oscillation experiments. In what regards inverted
neutrino mass hierarchy, we find from Eqs. (\ref{Rl}) and Eqs. (\ref{Rnu}),
that the corresponding PMNS leptonic mixing matrix takes the form: 
\begin{equation}
U=R^{\dagger}_lR_{\nu}=\left( 
\begin{array}{ccc}
\frac{1}{\sqrt{6}}\sin \theta _{l}e^{i\kappa }-\frac{\sqrt{6}}{3}\cos \theta
_{l} & \frac{1}{\sqrt{3}}\cos \theta _{l}+\frac{1}{\sqrt{3}}\sin \theta
_{l}e^{i\kappa } & \frac{1}{\sqrt{2}}\sin \theta _{l}e^{i\kappa } \\ 
\frac{1}{6}\sqrt{6} & \frac{1}{3}\sqrt{3} & -\frac{1}{\sqrt{2}} \\ 
\frac{1}{\sqrt{6}}\cos \theta _{l}+\frac{\sqrt{6}}{3}\sin \theta
_{l}e^{-i\kappa } & \frac{1}{\sqrt{3}}\cos \theta _{l}-\frac{1}{\sqrt{3}}%
\sin \theta _{l}e^{-i\kappa } & \frac{1}{\sqrt{2}}\cos \theta _{l}%
\end{array}%
\right) \allowbreak .  \label{UPMNS}
\end{equation}

From the standard parametrization of the leptonic mixing matrix, we predict
that the lepton mixing parameters for the case of inverted neutrino mass
hierarchy are given by:

\begin{eqnarray}
\sin ^{2}\theta _{12}&=&\frac{\left\vert U_{e2}\right\vert ^{2}}{%
1-\left\vert U_{e3}\right\vert ^{2}}=\frac{1+\cos \kappa \sin 2\theta _{l}}{3\left( 1-\frac{%
\sin ^{2}\theta _{l}}{2}\right) }\simeq 0.321,\notag\\
\sin ^{2}\theta_{23}&=&\frac{\left\vert U_{\mu 3}\right\vert ^{2}}{%
1-\left\vert U_{e3}\right\vert ^{2}}=\frac{1}{2\left( 1-\frac{\sin ^{2}\theta _{l}}{2}\right) }\simeq 0.511,\notag\\
\sin ^{2}\theta _{13}&=&\left\vert U_{e3}\right\vert ^{2}=\frac{\sin ^{2}\theta _{l}}{2}\simeq
0.0214.  \label{mixingparameters}
\end{eqnarray}
Let us note that for the sake of simplicity, we have taken reals the parameters $A_1$ and $A_2$ of the light active neutrino mass matrix. It is worth mentioning that the introduction of complex phases in the $A_1$ and $A_2$ parameters will not modify our predictions for the leptonic mixing parameters since they do not depend on the Majorana phases.

The obtained values for the charged lepton masses and leptonic mixing
parameters are obtained starting from the following benchmark point: 
\begin{equation}
x_{1}\simeq 1.524,\hspace{1cm}x_{2}\simeq 1.520,\hspace{1cm}y\simeq 1.025,%
\hspace{1cm}\left\vert z_{1}\right\vert \simeq 0.813,\hspace{1cm}z_{2}\simeq
0.864,\hspace{1cm}\kappa \simeq -81.82^{\circ }.  \label{fit}
\end{equation}

\begin{table}[tbh]
\begin{tabular}{|c|c|c|c|c|c|}
\hline
Parameter & $\Delta m_{21}^{2}$($10^{-5}$eV$^{2}$) & $\Delta m_{13}^{2}$($%
10^{-3}$eV$^{2}$) & $\left( \sin ^{2}\theta _{12}\right) _{\exp }$ & $\left(
\sin ^{2}\theta _{23}\right) _{\exp }$ & $\left( \sin ^{2}\theta
_{13}\right) _{\exp }$ \\ \hline
Best fit $\pm$ $1\sigma$ & $7.56\pm 0.19$ & $2.49\pm 0.04$ & $%
0.321_{-0.016}^{+0.018}$ & $0.596_{-0.018}^{+0.017}$ & $%
0.02140_{-0.00085}^{+0.00082}$ \\ \hline
$2\sigma $ range & $7.20-7.95$ & $2.41-2.57$ & $0.289-0.359$ & $0.404-0.456$
and $0.556-0.625$ & $0.0197-0.0230$ \\ \hline
$3\sigma $ range & $7.05-8.14$ & $2.37-2.61$ & $0.273-0.379$ & $0.388-0.638$
& $0.0189-0.0239$ \\ \hline
\end{tabular}%
\caption{Range for experimental values of neutrino mass squared splittings
and leptonic mixing parameters, taken from Ref. \protect\cite%
{deSalas:2017kay}, for the case of inverted hierarchy.}
\label{IH}
\end{table}

From the comparison of Eq. (\ref{mixingparameters}) with Table \ref{IH}, it
follows that the solar $\sin ^{2}\theta _{12}$ and reactor $\sin ^{2}\theta
_{13}$ mixing parameter are in excellent agreement with the experimental
data, whereas the atmospheric $\sin ^{2}\theta _{23}$ leptonic mixing
parameter is deviated $3\sigma $ away, respectively from its best fit value.
Let us note that with only two free effective parameters, i.e., $\theta _{l}$
and $\kappa $, our model predict leptonic mixing parameters in very good
agreement with their experimental values, for the case of inverted neutrino
mass spectrum. Furthermore, the obtained Jarlskog invariant and leptonic Dirac CP violating phase are given by: 
\begin{eqnarray}
J &=&\frac{1}{6}\sin \theta _{l}\cos \theta _{l}\sin \kappa \simeq
-3.34\times 10^{-2},  \notag \\
\delta  &=&\arcsin \left( \frac{\left( 3+\cos 2\theta _{l}\right) ^{\frac{3}{%
2}}\sin \kappa }{4\sqrt{\left( 4-4\cos \kappa \cos \theta _{l}\sin \theta
_{l}-3\sin ^{2}\theta _{l}\right) \left( 1+\cos \kappa \sin 2\theta
_{l}\right) }}\right) \simeq -81.37^{\circ }.  \label{J}
\end{eqnarray}

Furthermore, from the experimental values of the neutrino mass squared
splittings for the case of inverted neutrino mass hierarchy, we found that
the $A_{1\nu }$ and $A_{2\nu }$ parameters are given by:

\begin{equation*}
A_{1\nu }\simeq 0.0252\mbox{eV},\hspace{1cm}\hspace{1cm}A_{2\nu }\simeq
0.0489\mbox{eV}.
\end{equation*}

Thus, we obtain the following values for the neutrino mass squared
splittings for the case of inverted neutrino mass hierarchy:

\begin{equation}
\Delta m_{21}^{2}\simeq 7.56\times 10^{-5}\mbox{eV}^{2},\hspace{1cm}\hspace{%
1cm}\Delta m_{13}^{2}\simeq 2.49\times 10^{-3}\mbox{eV}^{2}.
\label{Deltamnusquad}
\end{equation}

Consequently, the predicted values for the neutrino mass squared splittings
are inside their $1\sigma $ experimentally allowed range, thus exhibiting an
excellent agreement with the experimental data on neutrino oscillations
experiments, as follows from the comparison of Eq. (\ref{Deltamnusquad})
with Table \ref{IH}.

\begin{figure}[tbp]
\includegraphics[width=0.5\textwidth]{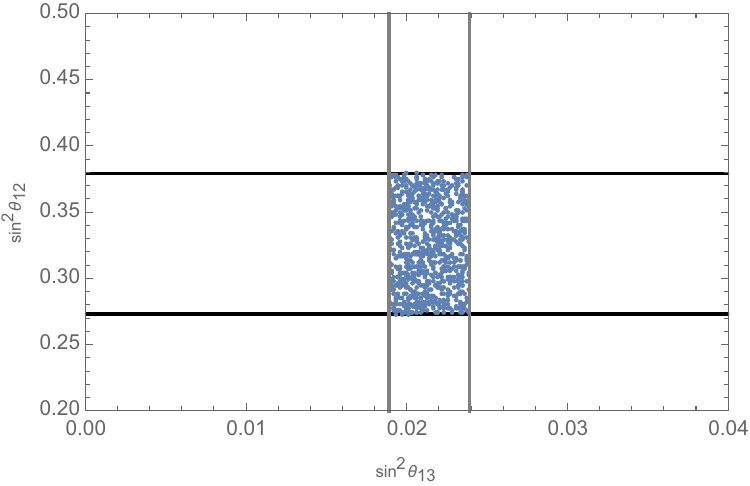}\includegraphics[width=0.5\textwidth]{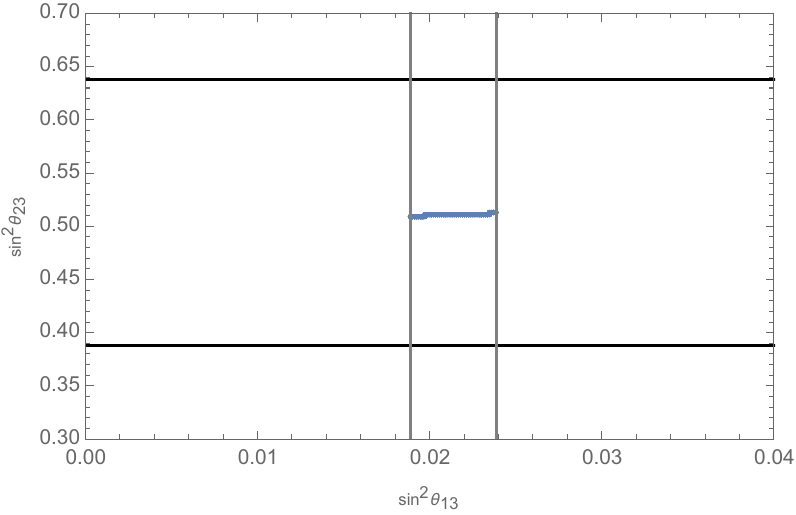}\newline
\caption{Correlations of the atmospheric $\sin ^{2}\protect\theta _{23}$ and
solar $\sin ^{2}\protect\theta _{12}$ mixing parameters with the reactor
mixing parameter $\sin^{2}\protect\theta _{13}$. The vertical lines are the
minimum and maximum values of the reactor mixing parameter $\sin^{2}\protect%
\theta _{13}$ inside the $3\protect\sigma$ experimentally allowed range.}
\label{Correlationsslij}
\end{figure}

\begin{figure}[tbp]
\includegraphics[width=0.5\textwidth]{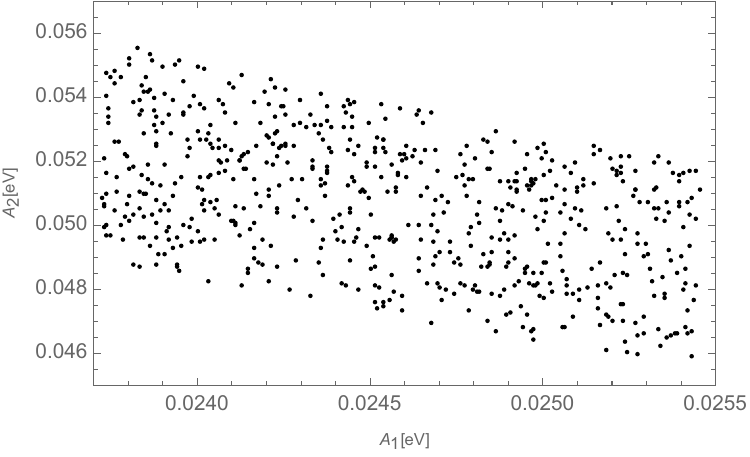}%
\includegraphics[width=0.5\textwidth]{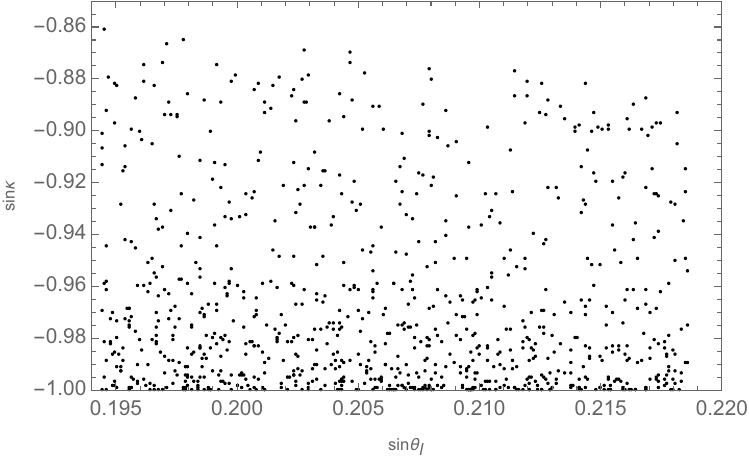}%
\newline
\caption{Correlations of the $A_1$ with $A_2$ mass parameters and $\sin%
\protect\theta_l$ and $\sin\protect\kappa$ mixing parameters that
sucessfully reproduce the values of the leptonic mixing parameters and
neutrino mass squared splittings inside the $3\protect\sigma$ experimentally
allowed range.}
\label{CorrelationsA1A2}
\end{figure}

\begin{figure}[tbp]
\vspace{-2cm} 
\resizebox{20cm}{30cm}{\hspace{0cm}
\includegraphics[width=0.5\textwidth]{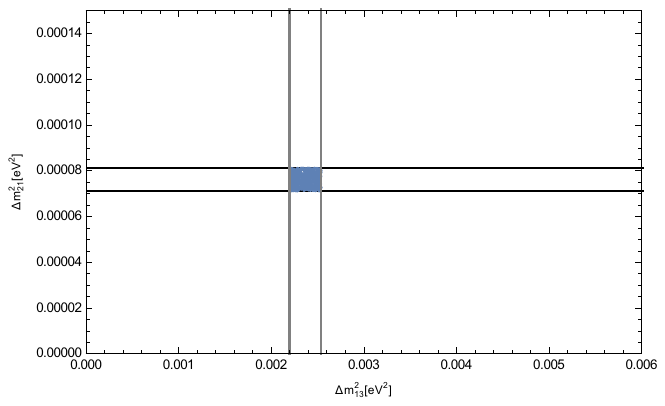}} \vspace{%
-20cm}
\caption{Correlations between $\Delta m_{21}^{2}$ and $\Delta m_{31}^{2}$.
The horizonal and vertical lines are the minimum and maximum values of the
neutrino mass squared splittings inside the $3\protect\sigma$ experimentally
allowed range.}
\label{Neutrinomasssplittings}
\end{figure}

Fig.~\ref{Correlationsslij} shows the correlations of the atmospheric $\sin ^{2}\protect\theta _{23}$ and
solar $\sin ^{2}\protect\theta _{12}$ mixing parameters with the reactor
mixing parameter $\sin^{2}\protect\theta _{13}$. For our analysis, we randomly generated
parameter configurations for $\theta_l$ and $\kappa$ corresponding to $3\sigma$ values for the leptonic mixing parameters. To this end, we varied the parameters $\theta_l$ and $\kappa$ in the ranges $11.21^{\circ}\leqslant\theta_l\leqslant 12.63^{\circ}$ and $-121.01^{\circ}\leqslant\kappa\leqslant -72.37^{\circ}$ (larger ranges will yields leptonic mixing parameters outside the $3\sigma$ experimentally allowed range). The correlation of the $A_1$ with $A_2$ mass parameters and $\sin\protect\theta_l$ and $\sin\protect\kappa$ mixing parameters that sucessfully reproduce the values of the leptonic mixing parameters and neutrino mass squared splittings inside the $3\protect\sigma$ experimentally allowed range are shown in Fig.~\ref{CorrelationsA1A2}. Correlations between the neutrino mass squared splittings $\Delta m_{21}^{2}$ and $\Delta m_{31}^{2}$ are displayed in Fig.~\ref{Neutrinomasssplittings}.

\begin{figure}[tbp]
\includegraphics[width=0.5\textwidth]{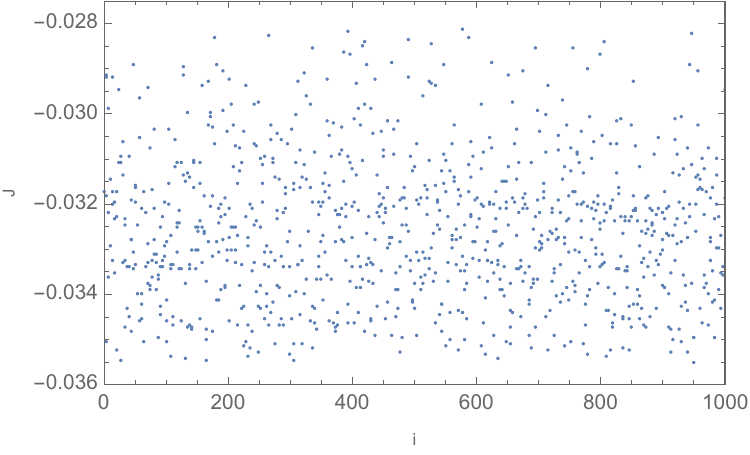}\includegraphics[width=0.5%
\textwidth]{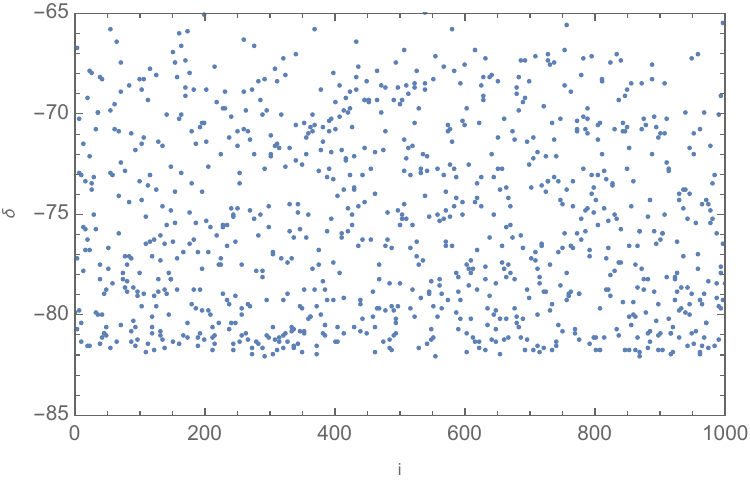}\newline
\caption{Jarlskog invariant and leptonic Dirac CP violating phase randomly
generated.}
\label{Jlanddeltal}
\end{figure}

Furthermore, in order to study the sensitivity of the obtained values for the Jarlskog invariant and leptonic Dirac CP violating phase under small variations around the best-fit values, subjected to the restriction that the resulting leptonic mixing parameters be inside the $3\sigma$ experimentally allowed range, we randomly generated parameter configurations for $\theta_l$ and $\kappa$ in the ranges $11.21^{\circ}\leqslant\theta_l\leqslant 12.63^{\circ}$ and $-121.01^{\circ}\leqslant\kappa\leqslant -72.37^{\circ}$, respectively. The resulting values for the Jarlskog invariant and leptonic Dirac CP violating phase as functions of the iteration are shown in Fig.~\ref{Jlanddeltal}. As indicated by Fig.~\ref{Jlanddeltal}, our model predicts Jarlskog invariant and leptonic Dirac CP violating phase in the ranges $-3.55\times 10^{-2}\leqslant J\leqslant -2.80\times 10^{-2}$ and $-82^{\circ}\leqslant\delta\leqslant -65^{\circ}$, respectively. 
%
%

%

In the following we proceed to determine the effective Majorana neutrino
mass parameter, whose value is proportional to the amplitude of neutrinoless
double beta ($0\nu \beta \beta $) decay. The effective Majorana neutrino
mass parameter is given by: 
\begin{equation}
m_{ee}=\left\vert \sum_{j}U_{ek}^{2}m_{\nu _{k}}\right\vert ,  \label{mee}
\end{equation}%
where $U_{ej}^{2}$ and $m_{\nu _{k}}$ are the squared of the PMNS leptonic
mixing matrix elements and the masses of the Majorana neutrinos,
respectively.

Thus, we obtain the following value for the effective Majorana neutrino mass
parameter in the case of inverted neutrino mass hierarchy: 
\begin{equation}
m_{ee}\simeq 46.9\mbox{meV}  \label{eff-mass-pred}
\end{equation}

In order to determine the predicted ranges for the effective Majorana neutrino mass parameter $m_{ee}$ in our model, we have randomly generated the parameters $A_1$, $A_2$, $\theta_l$ and $\kappa$ in a range of values where the neutrino mass squared splittings and the leptonic are consistent with the neutrino oscillation experimental data, in the scenario of vanishing Majorana phases. The effective Majorana neutrino mass parameter randomly generated as function of the iteration for the scenario of vanishing Majorana phases is shown in Fig.~\ref{mee}, which implies that this parameter has to be in the range $0.044$ eV$\lesssim m_{ee}\lesssim$ $0.048$ eV. In what regards, the scenario of nonvanishing Majorana phases, i.e., $A_1$ and $A_2$, complex, we have numerically checked that the obtained effective Majorana neutrino mass parameter has to be in the range $0.01$ eV $\lesssim m_{ee}\lesssim$ $0.05$ eV. 

\begin{figure}[tbp]
\includegraphics[width=0.5\textwidth]{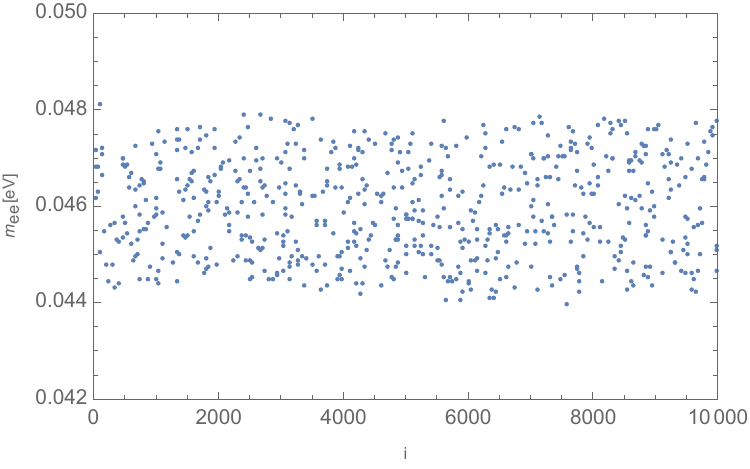}
\caption{Effective Majorana neutrino mass parameter randomly generated.}
\label{mee}
\end{figure}

Our obtained range of values for the effective Majorana neutrino mass parameter in the case of inverted neutrino mass
hierarchy, is within the declared reach of the next-generation bolometric
CUORE experiment \cite{Alessandria:2011rc} or, more realistically, of the
next-to-next-generation ton-scale $0\nu \beta \beta $-decay experiments. It
is worth mentioning that the effective Majorana neutrino mass parameter has
the upper bound of $m_{ee}\leq 160$ meV, which corresponds to $T_{1/2}^{0\nu
\beta \beta }(^{136}\mathrm{Xe})\geq 1.1\times 10^{26}$ yr at 90\% C.L,
which follows from the experimental data of the KamLAND-Zen experiment \cite%
{KamLAND-Zen:2016pfg}. That limit is expected to be updated in a not too
distant future. The GERDA \textquotedblleft phase-II\textquotedblright
experiment \cite{Abt:2004yk,Ackermann:2012xja} is expected to reach 
\mbox{$T^{0\nu\beta\beta}_{1/2}(^{76}{\rm Ge}) \geq
2\times 10^{26}$ yr}, which corresponds to $m_{ee}\leq 100$ meV. A
bolometric CUORE experiment, using ${}^{130}Te$ \cite{Alessandria:2011rc},
is currently under construction and its estimated sensitivity is about $%
T_{1/2}^{0\nu \beta \beta }(^{130}\mathrm{Te})\sim 10^{26}$ yr,
corresponding to \mbox{$m_{ee}\leq 50$ meV.} In addition, there are plans
for ton-scale next-to-next generation $0\nu \beta \beta $ experiments with $%
^{136}$Xe \cite{KamLANDZen:2012aa,Albert:2014fya} and $^{76}$Ge \cite%
{Abt:2004yk,Guiseppe:2011me}, asserting sensitivities over $T_{1/2}^{0\nu
\beta \beta }\sim 10^{27}$ yr, which corresponds to $m_{ee}\sim 12-30$ meV.
Some reviews on the theory and phenomenology of neutrinoless double-beta
decay are provided in Refs. \cite{Bilenky:2014uka,DellOro:2016tmg}. Our
results indicate that the derived model predicts $T_{1/2}^{0\nu \beta \beta
} $ at the level of sensitivities of the next generation or next-to-next
generation $0\nu \beta \beta $ experiments.

\section{Dark matter relic density.}

\label{DMsection} In this section we will discuss the implications of our
model in Dark matter. We will assume that the Dark matter candidate in the
model under consideration is a scalar. As a result of this assumption and
considering that the $SU\left( 3\right) _{L}$ scalar singlet $\varphi $ is
the only scalar field having a $Z_{6}$ charge corresponding to a nontrivial
charge under the preserved $Z_{2}$ symmetry, we have that either $\func{Im}%
\varphi $ or $\func{Re}\varphi $ can be a Dark matter candidate in our
model. Furthermore, we assume that the trilinear scalar coupling $f_{2}$
appearing in the scalar interaction $f_{2}\left[ \varrho \left( \varphi
^{\ast }\right) ^{2}+H.c\right] $ satisfies $f_{2}>0$, which implies that
the imaginary $\varphi _{I}=\func{Im}\varphi $ part of the scalar field $%
\varphi $ is lighter than its real part $\varphi _{R}=\func{Re}\varphi $, as
follows from Eq. (\ref{mphi}). Consequently $\func{Im}\varphi $ is the only
stable scalar field and thus the scalar Dark matter candidate in our model.

Relic density of the dark matter in the present Universe is estimated as
follows (c.f. Ref.~\cite{Olive:2016xmw}) 
\begin{equation}
\Omega h^{2}=\frac{0.1pb}{\left\langle \sigma v\right\rangle },\,\hspace{1cm}%
\left\langle \sigma v\right\rangle =\frac{A}{n_{eq}^{2}}\,,
\end{equation}%
where $\left\langle \sigma v\right\rangle $ is the thermally averaged
annihilation cross-section, $A$ is the total annihilation rate per unit
volume at temperature $T$ and $n_{eq}$ is the equilibrium value of the
particle density, which are given by \cite{Edsjo:1997bg} 
\begin{eqnarray}
A &=&\frac{T}{32\pi ^{4}}\dint\limits_{4m_{\varphi }^{2}}^{\infty
}\dsum\limits_{p=W,Z,t,b,h}g_{p}^{2}\frac{s\sqrt{s-4m_{\varphi }^{2}}}{2}%
v_{rel}\sigma \left( \varphi \varphi \rightarrow p\overline{p}\right)
K_{1}\left( \frac{\sqrt{s}}{T}\right) ds,  \notag \\
n_{eq} &=&\frac{T}{2\pi ^{2}}\dsum\limits_{p=W,Z,t,b,h}g_{p}m_{\varphi
}^{2}K_{2}\left( \frac{m_{\varphi }}{T}\right) ,
\end{eqnarray}%
with $K_{1}$ and $K_{2}$ being the modified Bessel functions of the second
kind order 1 and 2, respectively \cite{Edsjo:1997bg} and $m_{\varphi }=m_{%
\func{Im}\varphi }$. For the relic density calculation, we take $%
T=m_{\varphi}/20$ as in Ref. \cite{Edsjo:1997bg}, which corresponds to a
typical freeze-out temperature. We assume that our DM candidate $\varphi_I$
annihilates mainly into $WW$, $ZZ$, $t\overline{t}$, $b\overline{b}$ and $hh$%
, with annihilation cross sections given by: \cite{Bhattacharya:2016ysw}: 
\begin{eqnarray}
v_{rel}\sigma \left( \varphi _{I}\varphi _{I}\rightarrow WW\right) &=&\frac{%
\lambda _{h^{2}\varphi ^{2}}^{2}}{8\pi }\frac{s\left( 1+\frac{12m_{W}^{4}}{%
s^{2}}-\frac{4m_{W}^{2}}{s}\right) }{\left( s-m_{h}^{2}\right)
^{2}+m_{h}^{2}\Gamma _{h}^{2}}\sqrt{1-\frac{4m_{W}^{2}}{s}},  \notag \\
v_{rel}\sigma \left( \varphi _{I}\varphi _{I}\rightarrow ZZ\right) &=&\frac{%
\lambda _{h^{2}\varphi ^{2}}^{2}}{16\pi }\frac{s\left( 1+\frac{12m_{Z}^{4}}{%
s^{2}}-\frac{4m_{Z}^{2}}{s}\right) }{\left( s-m_{h}^{2}\right)
^{2}+m_{h}^{2}\Gamma _{h}^{2}}\sqrt{1-\frac{4m_{Z}^{2}}{s}},  \notag \\
v_{rel}\sigma \left( \varphi _{I}\varphi _{I}\rightarrow q\overline{q}%
\right) &=&\frac{N_{c}\lambda _{h^{2}\varphi ^{2}}^{2}m_{q}^{2}}{4\pi }\frac{%
\sqrt{\left( 1-\frac{4m_{f}^{2}}{s}\right) ^{3}}}{\left( s-m_{h}^{2}\right)
^{2}+m_{h}^{2}\Gamma _{h}^{2}},  \notag \\
v_{rel}\sigma \left( \varphi _{I}\varphi _{I}\rightarrow hh\right) &=&\frac{%
\lambda _{h^{2}\varphi ^{2}}^{2}}{16\pi s}\left( 1+\frac{3m_{h}^{2}}{%
s-m_{h}^{2}}-\frac{4\lambda _{h^{2}\varphi ^{2}}v^{2}}{s-2m_{h}^{2}}%
\right) ^{2}\sqrt{1-\frac{4m_{h}^{2}}{s}},
\end{eqnarray}%
where $\sqrt{s}$ is the centre-of-mass energy, $N_{c}=3$ is the color
factor, $m_{h}=125.7$ GeV and $\Gamma _{h}=4.1$ MeV are the SM Higgs boson $%
h $ mass and its total decay width, respectively.

In writting the above formulae we have considered that the scalar
interactions between the SM Higgs field $h$ and the scalar dark matter
candidate $\varphi $ are described by the following scalar potential:

\begin{equation}
V\left( \varphi ,h\right) =-\mu _{h}^{2}h^{2}+\frac{\mu _{\varphi }^{2}}{2}%
\left( \varphi _{R}^{2}+\varphi _{I}^{2}\right)-\frac{m_h^{2}}{2v}h^{3}+\frac{\lambda _{h^{4}}}{4!}%
h^{4}+\frac{\lambda _{\varphi ^{4}}}{4!}\left( \varphi _{R}^{2}+\varphi
_{I}^{2}\right) ^{2}+\frac{\lambda _{h^{2}\varphi ^{2}}}{4}\left( \varphi
_{R}^{2}+\varphi _{I}^{2}\right) h^{2}+\frac{\lambda _{h^{2}\varphi ^{2}}v}{2%
}\left( \varphi _{R}^{2}+\varphi _{I}^{2}\right) h\,.
\end{equation}
Here we have worked on the decoupling limit $\alpha-\beta=\frac{\pi}{2}$ where the couplings of the $126$ GeV Higgs boson to SM particles and its selfcouplings correspond to the SM expectation. 

Let us note that the tree level vacuum stability constraints resulting from
the requirement that the scalar potential be bounded from below, imply the
following relations \cite{EliasMiro:2012ay,Kannike:2016fmd}: 
\begin{equation}
\lambda _{h^{4}}>0,\,\hspace{1cm}\hspace{1cm}\lambda _{\varphi ^{4}}>0,\,%
\hspace{1cm}\hspace{1cm}\lambda _{h^{2}\varphi ^{2}}^{2}<\frac{2}{3}\lambda
_{h^{4}}\lambda _{\varphi ^{4}}.
\label{vs}
\end{equation}

Furthermore, the tree level unitarity constraints yields the following
relations \cite{Cynolter:2004cq}: 
\begin{equation}
\lambda _{\varphi ^{4}}<8\pi ,\,\hspace{1cm}\hspace{1cm}\lambda
_{h^{2}\varphi ^{2}}<4\pi.
\label{us}
\end{equation}

\begin{figure}[t]
\center
\vspace{0.8cm} \subfigure{\includegraphics[width=0.7%
\textwidth]{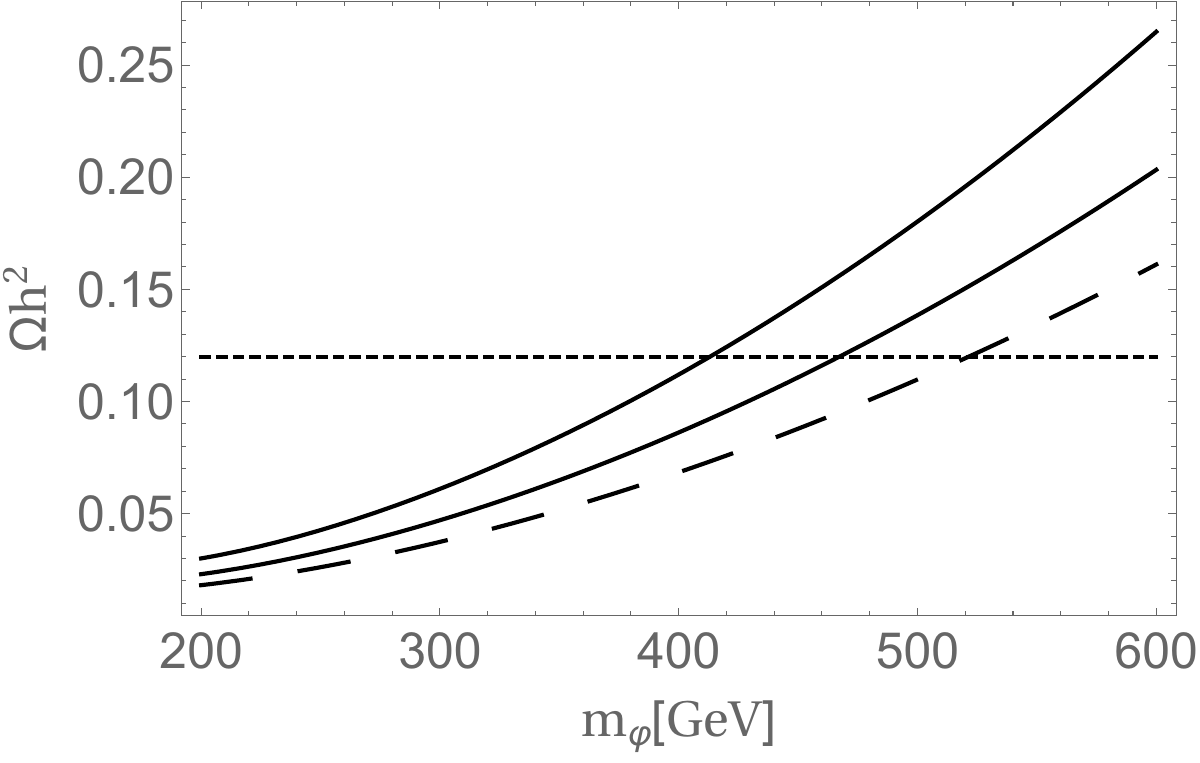}}
\caption{Relic density $\Omega h^{2}$, as a function of the mass $m_{\protect%
\varphi }$ of the $\protect\varphi $ scalar field, for several values of the
quartic scalar coupling $\protect\lambda _{h^{2}\protect\varphi ^{2}}$. The
curves from top to bottom correspond to $\protect\lambda _{h^{2}\protect%
\varphi ^{2}}=0.7,0.8,0.9$, respectively. The horizontal line shows the
observed value $\Omega h^{2}=0.1198$ \protect\cite{Ade:2015xua} for the
relic density.}
\label{DM}
\end{figure}

\begin{figure}[t]
\center
\vspace{0.8cm} \subfigure{\includegraphics[width=0.7%
\textwidth]{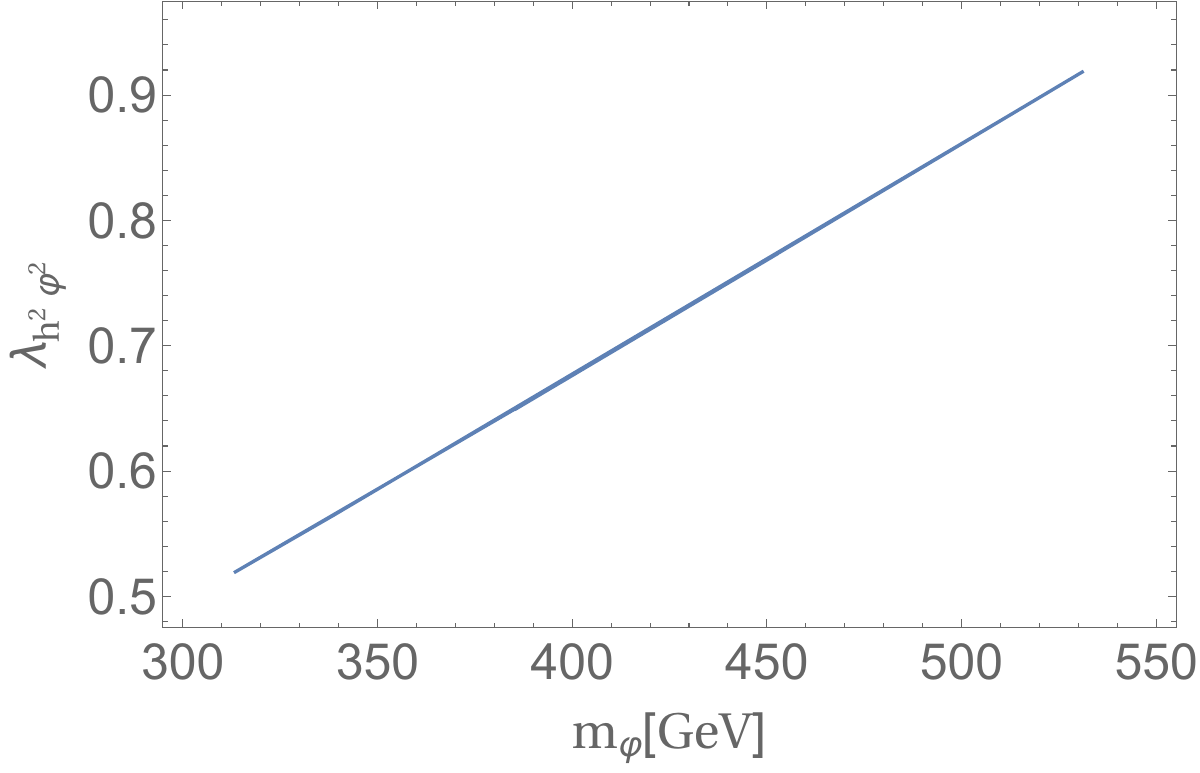}}
\caption{Correlation between the quartic scalar coupling and the mass $m_{%
\protect\varphi }$ of the scalar Dark matter candidate $\protect\varphi $,
consistent with the experimental value $\Omega h^{2}=0.1198$ for the Relic
density.}
\label{CorrelationDM}
\end{figure}

Fig.~\ref{DM} displays the Relic density $\Omega h^{2}$ as a function of the
mass $m_{\varphi }$ of the scalar field $\varphi _{I}$, for several values
of the quartic scalar coupling $\lambda _{h^{2}\varphi ^{2}}^{2}$. The
curves from top to bottom correspond to $\lambda _{h^{2}\varphi ^{2}}$ =0.7,
0.8 and 0.9, respectively. The horizontal line corresponds to the
experimental value $\Omega h^{2}=0.1198$ for the relic density. The Figure %
\ref{DM} shows that the Relic density is an increasing function of the mass $%
m_{\varphi }$ and a decreasing function of the quartic scalar coupling $%
\lambda _{h^{2}\varphi ^{2}}$. Consequently, an increase in the the mass $%
m_{\varphi }$ of the scalar field $\varphi _{I}$ will require a larger
quartic scalar coupling $\lambda _{h^{2}\varphi ^{2}}$, in order to account
for the measured value of the Dark matter relic density, as indicated by
Fig.~\ref{CorrelationDM}. It is worth mentioning that the Dark matter relic
density constraint yields a linear correlation between the quartic scalar
coupling $\lambda _{h^{2}\varphi ^{2}}$ and the mass $m_{\varphi }$ of the
scalar Dark matter candidate $\varphi _{I}$, as shown in Fig.~\ref%
{CorrelationDM}. We have numerically checked that in order to reproduce the
experimental value $\Omega h^{2}=0.1198\pm 0.0026$ \cite{Ade:2015xua} of the
relic density, the mass $m_{\varphi }$ of the scalar field $\varphi _{I}$\
has to be in the range $300$ GeV$\ \lesssim m_{\varphi }\lesssim $ $600$ GeV,%
$\ $for a quartic scalar coupling $\lambda _{h^{2}\varphi ^{2}}$ in the
range $0.5\lesssim \lambda _{h^{2}\varphi ^{2}}\lesssim 1$, which is consistent with the vacuum stability and unitarity constraints shown in Eqs. \ref{vs} and \ref{us}. Furthermore, our range of values chosen for the quartic scalar coupling $\lambda _{h^{2}\varphi ^{2}}$ also allow the extrapolation of our model at high energy scales as well as the preservation of perturbativity at one loop level.


\section{Conclusions}

\label{conclusions}

We constructed a highly predictive 3-3-1 model with right-handed neutrinos,
where the symmetry is extended by $A_{4}\times Z_{4}\times Z_{6}\times
Z_{16}\times Z_{16}^{\prime }$ and the field content is enlarged by extra $%
SU(3)_{L}$ singlet scalar fields and six right handed Majorana neutrinos.
Our model is consistent with the low energy fermion flavor data. The $A_{4}$%
, $Z_{4}$, $Z_{6}$, and $Z_{16}^{\prime }$ symmetries are crucial for
reducing the number of fermion sector model parameters, whereas the $Z_{16}$
symmetry causes the charged fermion mass and quark mixing pattern. In the
model under consideration, the light active neutrino masses are generated
from a one loop level inverse seesaw mechanism and the observed pattern of
charged fermion masses and quark mixing angles is caused by the breaking of
the $A_{4}\times Z_{4}\times Z_{6}\times Z_{16}\times Z_{16}^{\prime }$
discrete group at very high energy. In our model the different discrete
group factors are broken completely, excepting the $Z_{6}$ discrete group,
which is broken down to the preserved $Z_{2}$ symmetry, thus allowing the
implementation of the one loop level inverse seesaw mechanism for the
generation of the light active neutrino masses. The resulting the neutrino
spectrum of our model is composed of light active neutrinos and TeV scale
exotic pseudo-Dirac neutrinos. The smallness of the active neutrino masses
is a natural consequence of their scaling with inverse powers of the large
model cutoff $\Lambda $ and of their linear dependence on the loop induced
mass scale $m_{R}$ for the Majorana neutrinos $N_{i}$ ($i=1,2,3$). The
obtained values of the physical observables for the quark sector are
consistent with the experimental data, whereas the ones for the lepton
sector also do but only for the inverted neutrino mass spectrum. The normal
neutrino mass hierarchy scenario of our model is disfavored by the neutrino
oscillation experimental data, since the resulting reactor mixing parameter
is much larger than its experimental upper limit. The obtained model
predicts an effective Majorana neutrino mass parameter of neutrinoless
double beta decay of $m_{ee}=46.9$ meV, a leptonic Dirac CP violating phase
of $-81.37^{\circ }$ and a Jarlskog invariant of about $10^{-2}$ for the
inverted neutrino mass spectrum. Our obtained value of meV for the effective
Majorana neutrino mass is within the declared reach of the next generation
bolometric CUORE experiment \cite{Alessandria:2011rc} or, more
realistically, of the next-to-next generation ton-scale $0\nu \beta \beta $%
-decay experiments. Due to the fact that the $Z_{6}$ discrete group, which
is broken down to the preserved $Z_{2}$ symmetry our model possesses a
scalar DM particle candidate. The constraints arising from the DM relic
density, set its mass in the range $300$ GeV$\ \lesssim m_{\varphi }\lesssim 
$ $600$ GeV,$\ $for a quartic scalar coupling $\lambda _{h^{2}\varphi ^{2}}$
in the window $0.5\lesssim \lambda _{h^{2}\varphi ^{2}}\lesssim 1$. 

\section*{Acknowledgments}

This research has received funding from Fondecyt (Chile), Grants
No.~1170803, CONICYT PIA/Basal FB0821, the Vietnam National Foundation for
Science and Technology Development (NAFOSTED) under grant number
103.01-2017.356. A.E.C.H is very grateful to the Institute of Physics,
Vietnam Academy of Science and Technology for the warm hospitality and for
fully financing his visit.

\appendix

\section{The product rules for $A_4$}

\label{A4}The $A_{4}$ group has one three-dimensional $\mathbf{3}$\ and
three distinct one-dimensional $\mathbf{1}$, $\mathbf{1}^{\prime }$ and $%
\mathbf{1}^{\prime \prime }$ irreducible representations, satisfying the
following product rules: 
\begin{eqnarray}
&&\hspace{18mm}\mathbf{3}\otimes \mathbf{3}=\mathbf{3}_{s}\oplus \mathbf{3}%
_{a}\oplus \mathbf{1}\oplus \mathbf{1}^{\prime }\oplus \mathbf{1}^{\prime
\prime },  \label{A4-singlet-multiplication} \\[0.12in]
&&\mathbf{1}\otimes \mathbf{1}=\mathbf{1},\hspace{5mm}\mathbf{1}^{\prime
}\otimes \mathbf{1}^{\prime \prime }=\mathbf{1},\hspace{5mm}\mathbf{1}%
^{\prime }\otimes \mathbf{1}^{\prime }=\mathbf{1}^{\prime \prime },\hspace{%
5mm}\mathbf{1}^{\prime \prime }\otimes \mathbf{1}^{\prime \prime }=\mathbf{1}%
^{\prime },  \notag
\end{eqnarray}%
Considering $\left( x_{1},y_{1},z_{1}\right) $ and $\left(
x_{2},y_{2},z_{2}\right) $ as the basis vectors for two $A_{4}$-triplets $%
\mathbf{3}$, the following relations are fullfilled\textbf{:}

\begin{eqnarray}
&&\left( \mathbf{3}\otimes \mathbf{3}\right) _{\mathbf{1}%
}=x_{1}y_{1}+x_{2}y_{2}+x_{3}y_{3},  \label{triplet-vectors} \\
&&\left( \mathbf{3}\otimes \mathbf{3}\right) _{\mathbf{3}_{s}}=\left(
x_{2}y_{3}+x_{3}y_{2},x_{3}y_{1}+x_{1}y_{3},x_{1}y_{2}+x_{2}y_{1}\right) ,\
\ \ \ \left( \mathbf{3}\otimes \mathbf{3}\right) _{\mathbf{1}^{\prime
}}=x_{1}y_{1}+\omega x_{2}y_{2}+\omega ^{2}x_{3}y_{3},  \notag \\
&&\left( \mathbf{3}\otimes \mathbf{3}\right) _{\mathbf{3}_{a}}=\left(
x_{2}y_{3}-x_{3}y_{2},x_{3}y_{1}-x_{1}y_{3},x_{1}y_{2}-x_{2}y_{1}\right) ,\
\ \ \left( \mathbf{3}\otimes \mathbf{3}\right) _{\mathbf{1}^{\prime \prime
}}=x_{1}y_{1}+\omega ^{2}x_{2}y_{2}+\omega x_{3}y_{3},  \notag
\end{eqnarray}%
where $\omega =e^{i\frac{2\pi }{3}}$. The representation $\mathbf{1}$ is
trivial, while the non-trivial $\mathbf{1}^{\prime }$ and $\mathbf{1}%
^{\prime \prime }$ are complex conjugate to each other. Some reviews of
discrete symmetries in particle physics are found in Refs. \cite%
{Ishimori:2010au,Altarelli:2010gt,King:2013eh, King:2014nza}.

\section{Scalar potential for one $A_{4}$ scalar triplet}

\label{B1}

The scalar potential for any $A_{4}$ scalar triplet takes the form: 
\begin{eqnarray}
V\left( \Sigma \right) &=&-\mu _{\Sigma }^{2}\left( \Sigma \Sigma ^{\ast
}\right) _{\mathbf{1}}+\kappa _{\Sigma ,1}\left( \Sigma \Sigma ^{\ast
}\right) _{\mathbf{1}}\left( \Sigma \Sigma ^{\ast }\right) _{\mathbf{1}%
}+\kappa _{\Sigma ,2}\left( \Sigma \Sigma \right) _{\mathbf{1}}\left( \Sigma
^{\ast }\Sigma ^{\ast }\right) _{\mathbf{1}}+\kappa _{\Sigma ,3}\left(
\Sigma \Sigma ^{\ast }\right) _{\mathbf{1}^{\prime }}\left( \Sigma \Sigma
^{\ast }\right) _{\mathbf{1}^{\prime \prime }}  \notag \\
&&+\kappa _{\Sigma ,4}\left[ \left( \Sigma \Sigma \right) _{\mathbf{1}%
^{\prime }}\left( \Sigma ^{\ast }\Sigma ^{\ast }\right) _{\mathbf{1}^{\prime
\prime }}+h.c\right] +\kappa _{\Sigma ,5}\left[ \left( \Sigma \Sigma \right)
_{\mathbf{1}^{\prime \prime }}\left( \Sigma ^{\ast }\Sigma ^{\ast }\right) _{%
\mathbf{1}^{\prime }}+h.c\right]  \notag \\
&&+\kappa _{\Sigma ,6}\left( \Sigma \Sigma ^{\ast }\right) _{\mathbf{3s}%
}\left( \Sigma \Sigma ^{\ast }\right) _{\mathbf{3s}}+\kappa _{\Sigma
,7}\left( \Sigma \Sigma \right) _{\mathbf{3s}}\left( \Sigma ^{\ast }\Sigma
^{\ast }\right) _{\mathbf{3s}}.  \label{ScalarpotentialA4triplet}
\end{eqnarray}%
where $\Sigma =\xi $, $\zeta $, $\Phi $, $\Delta $, $\Xi $, $\Theta $.

That scalar potential given above has 8 free parameters: 1 bilinear and 7
quartic couplings. The scalar potential minimization conditions read: 
\begin{eqnarray}
\frac{\partial \left\langle V\left( \Sigma \right) \right\rangle }{\partial 
\text{$v_{\Sigma _{1}}$}} &=&-2v_{\Sigma _{1}}\mu _{\Sigma }^{2}+4\kappa
_{\Sigma ,1}\text{$v_{\Sigma _{1}}$}\left( \text{$v_{\Sigma
_{1}}^{2}+v_{\Sigma _{2}}^{2}+v_{\Sigma _{3}}^{2}$}\right) +2\kappa _{\Sigma
,3}\text{$v_{\Sigma _{1}}$}\left( 2\text{$v_{\Sigma _{1}}^{2}-v_{\Sigma
_{2}}^{2}-v_{\Sigma _{3}}^{2}$}\right)  \notag \\
&&+4\kappa _{\Sigma ,2}\text{$v_{\Sigma _{1}}$}\left[ \text{$v_{\Sigma
_{1}}^{2}+v_{\Sigma _{2}}^{2}\cos \left( 2\theta _{\Sigma _{1}}-2\theta
_{\Sigma _{2}}\right) +v_{\Sigma _{3}}^{2}\cos \left( 2\theta _{\Sigma
_{1}}-2\theta _{\Sigma _{3}}\right) $}\right] +8\kappa _{\Sigma ,7}\text{$%
v_{\Sigma _{1}}$}\left( v_{\Sigma _{2}}^{2}+v_{\Sigma _{3}}^{2}\right) 
\notag \\
&&+4\left( \kappa _{\Sigma ,4}+\kappa _{\Sigma ,5}\right) \text{$v_{\Sigma
_{1}}$}\left[ 2\text{$v_{\Sigma _{1}}^{2}-v_{\Sigma _{2}}^{2}\cos \left(
2\theta _{\Sigma _{1}}-2\theta _{\Sigma _{2}}\right) -v_{\Sigma
_{3}}^{2}\cos \left( 2\theta _{\Sigma _{1}}-2\theta _{\Sigma _{3}}\right) $}%
\right]  \notag \\
&&+4\kappa _{\Sigma ,6}\text{$v_{\Sigma _{1}}$}\left[ \text{$v_{\Sigma
_{2}}^{2}\left\{ 1+\cos \left( 2\theta _{\Sigma _{1}}-2\theta _{\Sigma
_{2}}\right) \right\} +v_{\Sigma _{3}}^{2}\left\{ 1+\text{$\cos \left(
2\theta _{\Sigma _{1}}-2\theta _{\Sigma _{3}}\right) $}\right\} $}\right] 
\notag \\
&=&0,  \notag \\
\frac{\partial \left\langle V\left( \Sigma \right) \right\rangle }{\partial 
\text{$v_{\Sigma _{2}}$}} &=&-2v_{\Sigma _{2}}\mu _{\Sigma }^{2}+4\kappa
_{\Sigma ,1}\text{$v_{\Sigma _{2}}$}\left( \text{$v_{\Sigma
_{1}}^{2}+v_{\Sigma _{2}}^{2}+v_{\Sigma _{3}}^{2}$}\right) +2\kappa _{\Sigma
,3}\text{$v_{\Sigma _{2}}$}\left( 2\text{$v_{\Sigma _{2}}^{2}-v_{\Sigma
_{1}}^{2}-v_{\Sigma _{3}}^{2}$}\right)  \notag \\
&&+4\kappa _{\Sigma ,2}\text{$v_{\Sigma _{2}}$}\left[ v_{\Sigma _{2}}^{2}+%
\text{$v_{\Sigma _{1}}^{2}\cos \left( 2\theta _{\Sigma _{2}}-2\theta
_{\Sigma _{1}}\right) +v_{\Sigma _{3}}^{2}\cos \left( 2\theta _{\Sigma
_{2}}-2\theta _{\Sigma _{3}}\right) $}\right] +8\kappa _{\Sigma ,7}\text{$%
v_{\Sigma _{2}}$}\left( v_{\Sigma _{1}}^{2}+v_{\Sigma _{3}}^{2}\right) 
\notag \\
&&+4\left( \kappa _{\Sigma ,4}+\kappa _{\Sigma ,5}\right) \text{$v_{\Sigma
_{2}}$}\left[ 2v_{\Sigma _{2}}^{2}-\text{$v_{\Sigma _{1}}^{2}\cos \left(
2\theta _{\Sigma _{2}}-2\theta _{\Sigma _{1}}\right) -v_{\Sigma
_{3}}^{2}\cos \left( 2\theta _{\Sigma _{2}}-2\theta _{\Sigma _{3}}\right) $}%
\right]  \notag \\
&&+4\kappa _{\Sigma ,6}\text{$v_{\Sigma _{2}}$}\left[ \text{$v_{\Sigma
_{1}}^{2}\left\{ 1+\cos \left( 2\theta _{\Sigma _{2}}-2\theta _{\Sigma
_{1}}\right) \right\} +v_{\Sigma _{3}}^{2}\left\{ 1+\text{$\cos \left(
2\theta _{\Sigma _{2}}-2\theta _{\Sigma _{3}}\right) $}\right\} $}\right] 
\notag \\
&=&0,  \notag \\
\frac{\partial \left\langle V\left( \Sigma \right) \right\rangle }{\partial 
\text{$v_{\Sigma _{3}}$}} &=&-2v_{\Sigma _{3}}\mu _{\Sigma }^{2}+4\kappa
_{\Sigma ,1}\text{$v_{\Sigma _{3}}$}\left( \text{$v_{\Sigma
_{1}}^{2}+v_{\Sigma _{2}}^{2}+v_{\Sigma _{3}}^{2}$}\right) +2\kappa _{\Sigma
,3}\text{$v_{\Sigma _{3}}$}\left( 2\text{$v_{\Sigma _{3}}^{2}-v_{\Sigma
_{1}}^{2}-v_{\Sigma _{2}}^{2}$}\right)  \notag \\
&&+4\kappa _{\Sigma ,2}\text{$v_{\Sigma _{3}}$}\left[ v_{\Sigma _{2}}^{2}+%
\text{$v_{\Sigma _{1}}^{2}\cos \left( 2\theta _{\Sigma _{1}}-2\theta
_{\Sigma _{2}}\right) +v_{\Sigma _{2}}^{2}\cos \left( 2\theta _{\Sigma
_{3}}-2\theta _{\Sigma _{2}}\right) $}\right] +8\kappa _{\Sigma ,7}\text{$%
v_{\Sigma _{3}}$}\left( v_{\Sigma _{1}}^{2}+v_{\Sigma _{2}}^{2}\right) 
\notag \\
&&\text{$+$}4\left( \kappa _{\Sigma ,4}+\kappa _{\Sigma ,5}\right) \text{$%
v_{\Sigma _{3}}$}\left[ 2v_{\Sigma _{3}}^{2}-\text{$v_{\Sigma _{1}}^{2}\cos
\left( 2\theta _{\Sigma _{1}}-2\theta _{\Sigma _{2}}\right) -v_{\Sigma
_{2}}^{2}\cos \left( 2\theta _{\Sigma _{3}}-2\theta _{\Sigma _{2}}\right) $}%
\right]  \notag \\
&&+4\kappa _{\Sigma ,6}\text{$v_{\Sigma _{3}}$}\left[ \text{$v_{\Sigma
_{1}}^{2}\left\{ 1+\cos \left( 2\theta _{\Sigma _{1}}-2\theta _{\Sigma
_{2}}\right) \right\} +v_{\Sigma _{2}}^{2}\left\{ 1+\text{$\cos \left(
2\theta _{\Sigma _{3}}-2\theta _{\Sigma _{2}}\right) $}\right\} $}\right] 
\notag \\
&=&0.  \label{DV}
\end{eqnarray}%
where $\left\langle \Sigma \right\rangle =\left( \text{$v_{\Sigma
_{1}}e^{i\theta _{\Sigma _{1}}},v_{\Sigma _{2}}e^{i\theta _{\Sigma
_{2}}},v_{\Sigma _{3}}e^{i\theta _{\Sigma _{3}}}$}\right) $. Here for the
sake of simplicity we consider vanishing phases in the VEV patterns of the $%
A_{4}$ triplet scalars, i.e., $\theta _{\Sigma _{1}}=\theta _{\Sigma
_{2}}=\theta _{\Sigma _{3}}=0$. Then, the scalar potential minimization
equations given by Eq. (\ref{DV}) yields the following relations: 
\begin{eqnarray}
\left[ 3\kappa _{\Sigma ,3}-4\left( \kappa _{\Sigma ,6}+\kappa _{\Sigma
,7}\right) +6\left( \kappa _{\Sigma ,4}+\kappa _{\Sigma ,5}\right) \right]
\left( \text{$v_{\Sigma _{1}}^{2}-v_{\Sigma _{2}}^{2}$}\right) &=&0,  \notag
\\
\left[ 3\kappa _{\Sigma ,3}-4\left( \kappa _{\Sigma ,6}+\kappa _{\Sigma
,7}\right) +6\left( \kappa _{\Sigma ,4}+\kappa _{\Sigma ,5}\right) \right]
\left( \text{$v_{\Sigma _{1}}^{2}-v_{\Sigma _{3}}^{2}$}\right) &=&0,  \notag
\\
\left[ 3\kappa _{\Sigma ,3}-4\left( \kappa _{\Sigma ,6}+\kappa _{\Sigma
,7}\right) +6\left( \kappa _{\Sigma ,4}+\kappa _{\Sigma ,5}\right) \right]
\left( \text{$v_{\Sigma _{2}}^{2}-v_{\Sigma _{3}}^{2}$}\right) &=&0.
\label{DVb}
\end{eqnarray}%
From the relations given by Eq. (\ref{DVb}) and setting $\kappa _{S,3}=\frac{%
4}{3}\left( \kappa _{S,6}+\kappa _{S,7}\right) -2\left( \kappa _{S,4}+\kappa
_{S,5}\right) $, with $S=$ $\zeta $, $\Phi $, $\Delta $, $\Xi $, $\Theta $,
we obtain that the following VEV pattern: 
\begin{eqnarray}
\left\langle \xi \right\rangle &=&\frac{v_{\xi }}{\sqrt{3}}\left(
1,1,1\right) ,\hspace{1cm}\left\langle \Phi \right\rangle =v_{\Phi }\left(
1,0,0\right) ,\hspace{1cm}\left\langle \Delta \right\rangle =v_{\Delta
}\left( 0,0,1\right) ,  \notag \\
\left\langle \Xi \right\rangle &=&v_{\Xi }\left( 0,1,0\right) ,\hspace{1cm}%
\left\langle \zeta \right\rangle =\frac{v_{\zeta }}{\sqrt{2}}\left(
0,-1,1\right) ,\hspace{1cm}\left\langle \Theta \right\rangle =-\frac{%
v_{\Theta }}{\sqrt{5}}\left( 1,2,0\right).
\label{VEVpatternA4triplets}
\end{eqnarray}

is a solution of the scalar potential minimization equations for a large
region of parameter space.

From the expressions given above, and using the vacuum configuration for the 
$A_{4}$ scalar triplets given in Eq. (\ref{VEVpattern}), we find the following
relation: 
\begin{equation}
\mu _{\Sigma }^{2}=\frac{2}{3}\left[ 3\left( \kappa _{\Sigma ,1}+\kappa
_{\Sigma ,2}\right) +4\left( \kappa _{\Sigma ,6}+\kappa _{\Sigma ,7}\right) %
\right] v_{\Sigma }^{2},\hspace{1cm}\Sigma =\xi ,\Phi ,\Delta ,\Xi ,\Theta .
\end{equation}

These results indicate that the VEV patterns of the $A_{4}$ triplets, i.e., $%
\xi $, $\zeta $, $\Phi $, $\Delta $, $\Xi $ and $\Theta $ in Eq. (\ref%
{VEVpatternA4triplets}), are consistent with a global minimum of the scalar potential (%
\ref{ScalarpotentialA4triplet}) of our model for a large region of parameter
space.

\end{document}